\documentclass[12pt]{article}

\usepackage[T1]{fontenc}
\usepackage{latexsym}
\usepackage{physics}
\usepackage{amsmath}
\usepackage{scrextend}

\usepackage{amssymb}
\usepackage{slashed}
\usepackage{empheq}
\usepackage[english]{babel}
\usepackage{booktabs} 
\usepackage{array}
\usepackage{verbatim} 
\usepackage{xcolor}
\usepackage{soul}
\usepackage[pdfa]{hyperref}
\usepackage[]{pdfpages}
\usepackage{graphicx}

\begin{document}
\title{Tales from the prehistory of Quantum Gravity. L\'eon Rosenfeld's earliest contributions.}

\author{Giulio Peruzzi\thanks{{giulio.peruzzi@unipd.it, Department of Physics and Astronomy ``G. Galilei'', via Marzolo 8, I-35131 Padova (Italy)}} 
\and Alessio Rocci\thanks{{a\_rocci@hotmail.com, Department of Physics and Astronomy ``G. Galilei'', via Marzolo 8, I-35131 Padova (Italy)}}}

 \maketitle

\section*{Abstract}

The main purpose of this paper is to analyse the earliest work of L\'eon Rosenfeld, one of the pioneers in the search of Quantum Gravity, the supposed theory unifying quantum theory and general relativity. We describe how and why Rosenfeld tried to face this problem in 1927, analysing the role of his mentors: Oskar Klein, Louis de Broglie and Th\'eophile De Donder. Rosenfeld asked himself how quantum mechanics should \textit{concretely} modify general relativity. In the context of a five-dimensional theory, Rosenfeld tried to construct a unifying framework for the gravitational and electromagnetic interaction and wave mechanics. Using a sort of ``general relativistic quantum mechanics'' Rosenfeld introduced a wave equation on a curved background. He investigated the metric created by what he called `quantum phenomena', represented by wave functions. Rosenfeld integrated Einstein equations in the weak field limit, with wave functions as source of the gravitational field. The author performed a sort of semi-classical approximation obtaining at the first order the Reissner-Nordstr\"om metric. We analyse how Rosenfeld's work is part of the history of Quantum Mechanics, because in his investigation Rosenfeld was guided by Bohr's correspondence principle. Finally we briefly discuss how his contribution is connected with the task of finding out which metric can be generated by a quantum field, a problem that quantum field theory on curved backgrounds will start to address 35 years later.

\begin{quotation}\small\selectlanguage{english}
`A study of history of science [...] shows that the natural attitude of a scientist is to be inspired by their predecessors, but always taking the liberty of doubting when there are reasons for doubt.'
\begin{flushright}
\textit{Oskar Klein}
\end{flushright}
\end{quotation}

\tableofcontents

\section{Introduction}\label{intro}
In the physics community, the word quantum gravity (QG) is today associated with the task of quantizing gravity, directly or indirectly, in order to unravel a quantum structure of space and time. Despite many approaches, e.g. String Theory, Supergravity ($ N=8 $), Loop Quantum Gravity, non-commutative geometry and so on, a consistent theory is still lacking. From the point of view of History and Philosophy of Science: `QG, \textit{broadly construed}, is the physical theory (still ``under construction'') incorporating both the principles of general relativity (GR) and quantum theory' [emphasis added] (\cite{Stanford}). ``Broadly construed'' means that all the attempts  in this direction have contributed to our modern understanding of the difficulties in constructing a consistent theory of QG, even those approaches that did not quantize the gravitational interaction. To name one, quantum field theory (QFT) on curved backgrounds increased our knowledge on the physics of Black Holes \cite{Hawking}. Furthermore, from a point of view of the integrated History and Philosophy of Science (\&HPS), the fact that the theory is still under construction represents a unique opportunity for studying the process of a theory's formation from the inside (in Kuhnian words ``a revolution in progress'').

Usually the history of QG starts in 1930 with the first attempts to reconcile the budding quantum field theory with gravity made by L\'eon Rosenfeld \cite{Rosenfeld1} \cite{Rosenfeld2} (cf. English translation \cite{engl-transl} and the accompanying commentary \cite{commentary}). In the first paper the author tried to find out what would be the gravitational field produced by light in a weak-field approximation. This paper marked the beginning of what is today called the \textit{covariant approach}. In this work the quantization procedure was applied to the electromagnetic field only, the metric field being an operator because it is a function of the Maxwell field. In the second paper, conversely, he tried to apply the quantization procedure directly to the gravitational interaction, employing a tetrad gravitational field rather than the conventional metric. This paper marked the beginning of the today called \textit{canonical approach}. Before Rosenfeld's attempts, soon after the birth of GR in 1915, researchers tried to apply the theory of gravity to the microscopic world. The best known example is Einstein's claim of 1916. When he discovered that a mass should emit gravitational waves, Einstein pointed out the need to modify GR \cite{Ein1}. Of course what he had in mind was Bohr's old move that classical electrodynamics was not applicable in his model of orbiting electrons. In a similar way GR had to be modified with respect to its application to the microscopic world. Einstein's suggestion was not an isolated episode. Recent developments in the history of QG show that in the fifteen years before Rosenfeld's attempts many authors tried to reconcile the old quantum theory or quantum mechanics (QM) with gravity \cite{Stachel1} \cite{Rickles-presentation} \cite{Rickles-sources} \cite{Hagar} \cite{Rocci-Lodge} \cite{Rocci-tesi}. For this reason the period between 1915 and 1930 could be called a prehistory era. 

Exploring this time frame, the term ``quantum gravity'' must be necessarily interpreted in a broad sense, because in the period between 1916 and 1930 the quantization procedure was a concept under construction. As far as we know, before 1930, there were no attempts that tried to quantize the gravitational field directly. Before going on, we therefore briefly summarize the evolution of the quantization procedure during this period \cite{history-QM1}. Between 1916 and 1924, the construction of atomic models was one of the main tasks of the old quantum theory. The quantization procedure of the atomic model was performed by applying the Epstein-Sommerfeld-Wilson rules. After 1925, with the birth of QM, the investigation of the atomic phenomena was pursued by wave mechanics (WM) and matrix mechanics (MM). In the first formulation of QM, electrons are represented by normalized wave functions. WM was born by using Hamilton Jacobi (HJ) analogy between particle and waves \cite{Schroe1a}. The quantization procedure consisted in writing a wave equation and in imposing the boundary condition on wave functions. The second formulation of QM focused on observable quantities. MM was born by attempting to formulate a new theoretical technique for the determination of the intensities of quantum transitions, using the anharmonic oscillator as a toy model \cite{Blum2017}. The classical position and its conjugated momentum in the Hamiltonian formulation were treated as ``q-numbers'', that today are known as operators. The name ``q-numbers'' stands for quantum numbers, in contrast with ``c-numbers'', i.e. the usual classical variables, like e.g. classical position and momentum of a particle \cite{Darrigol}. The quantization procedure consisted in imposing the commutation relations between these q-numbers. In 1926 Schr\"odinger pointed out the equivalence between the two formulations, but WM remained the preferred point of view in attempting to generalize Schr\"odinger approach in the context of both Special and General Relativity \cite{Rocci-tesi}. In 1927 many new concepts were introduced: the description of spin with two components wave functions, its statistical interpretation, the uncertainty relations. At the end of 1927 Oskar Klein and Pascual Jordan introduced for the first time the quantum commutation relations for the scalar field operators, but the general approach was developed by Heisenberg and Pauli at the end of 1929. 

Rosenfeld was a protagonist of this early period as well. As stated in the introduction of a recent biography of Rosenfeld \cite{Rosenfeld-bio}, the Belgian physicist is a blank sheet in the history of science literature, `but he was at the centre of modern physics as one of the pioneers of quantum field theory and quantum
electrodynamics in the late 1920s and the 1930s' (\cite{Rosenfeld-bio}; p. 1). In spite of the fact that he initiated two of the major research areas in the history of QG, the covariant and the canonical approaches, Rosenfeld never considered his early work as an important contribution \cite{Kuhn1}. The aim of this paper is to offer a historical analysis ``in context'' of the papers published by Rosenfeld at the beginning of his career: \cite{Ros1}, \cite{Ros2},\cite{Ros3}, \cite{Ros4}, \cite{Ros5}. In particular we will focus on the aspects concerning the conciliation between GR and the WM, that produced a first attempt to find the metric generated by ``charged quantum matter'', using a wave-mechanical approach. Rosenfeld was persuaded, at that time, that he had found a quantum modification of the flat metric, using the correspondence principle. He performed a semi-classical approximation in order to compare his quantum metric with the external Reissner-Nordstr\"om (RN) metric. Aside from the fact that this attempt is important by itself, it contained the seeds for his following work \cite{Rosenfeld1}, nevertheless Rosenfeld later become one of the opponents to any quantization of the gravitational field without any experimental evidence for the necessity to do it \cite{Rosenfeld63}.

The paper is organized as follows. In section \ref{prehist} we briefly introduce Rosenfeld's life and we put it in the context of the prehistory of QG. In section \ref{5D} we review the work of the authors that influenced the professional training of the young Rosenfeld in 1927: Oskar Klein, Louis de Broglie and Th\'eophile De Donder. In particular we will focus on the analogies and on the differences among these authors. In section \ref{Rosenfeld} we present Rosenfeld's attempt to reconcile GR with WM. At the beginning we shall focus on his first paper, discussing how Klein, de Broglie and De Donder influenced Rosenfeld's work. Then we shall review the papers written by Rosenfeld in 1927, where a general relativistic version of Bohr's correspondence principle emerged. We shall also analyse the role played by Klein, and indirectly by Bohr, in suggesting the first use of the correspondence principle in the context of QG. At the beginning of section \ref{Rosenfeld} we shall focus on what Rosenfeld wanted to achieve. In the last part of the section, i.e. \ref{discussion}, we briefly present a modern interpretation of his approach and a perspective on how the analysed papers would influence Rosenfeld's subsequent work on the search of a quantum theory of gravity. In section \ref{summary} we summarize the basic stages of our paper without entering into technical details.

In the Appendices, section \ref{apps}, we describe with more details some calculations left out in the main text.

\section{The prehistory of QG and the young Rosenfeld}\label{prehist}
The prehistory of QG can be naturally divided into two parts. The first period from 1915 to 1924, was dominated by attempts to understand the role of GR in constructing planetary models of atoms \cite{Jaffe} \cite{Jeffery} \cite{Lodge} \cite{Vallarta1}. With the birth of QM in 1925-26 a new era began, because the classical concept of trajectory had become problematic in the atomic realm. In particular, the second period of the prehistory of QG from 1925 to 1930, was dominated by WM and by attempts which tried to generalize Schr\"odinger's approach in the context of Special Relativity (SR) and GR. In fact, between the two alternative formulations of QM, MM and WM, the second formulation was the preferred one by the authors of the period who tried to find a unique framework describing quantum phenomena and the gravitational interaction \cite{Rocci-tesi}. In this respect, as we will see, L\'eon Rosenfeld was not an exception. 

The career of the young Belgian physicist had started with the accidental reading of Schr\"odinger's communications \cite{Schroe1a}, as he recollected during an interview with Thomas S. Kuhn and John L. Heilbron in 1963 \cite{Kuhn1}. After completing his studies, Rosenfeld left the University of Li\`ege and moved to Paris at the end of 1926 to meet Louis de Broglie, where, as he recollected in the interview, he spent most of his time learning what he had missed at Li\`ege \cite{Kuhn1}. Rosenfeld himself stressed that he attended a course on relativity in Li\`ege and that the lecturer was an opponent of the new theory. In Paris, he attended many lectures, e.g. Langevin's lectures at the College de France, and he studied a lot of books, including Eddington's book on GR \cite{Eddington}: `I was anxious to do some research, and then the only research I did was in just combining my freshly acquired knowledge of relativity with wave mechanics [...]' \cite{Kuhn1}.

A key ingredient of this second period in the prehistory of QG is the enlargement of the four-dimensional space-time by the introduction of a fifth space-like dimension in order to look for a unified picture of the gravitational force, the electromagnetic interaction and the quantum behaviour of particles, described by a wave function. The idea was not new. The founding father of this approach is Theodore Kaluza \cite{Kaluza} who had noted that a five-dimensional theory of ``pure gravity'', i.e. without any matter content but with the electromagnetic potentials represented by specific components of the metric field, seems to offer a unified framework to describe the usual four-dimensional gravitational and electromagnetic interactions\footnote{More precisely Gunnar Nordstr\"om also tried a similar approach before Kaluza \cite{Nord}, but the Norwegian mathematician described the gravitational interaction using a scalar field instead of a tensor field.}. In 1927 many authors tried to harmonize Kaluza's picture with WM, and started explicitly from the German physicist's 1921 paper\footnote{Kaluza's approach was completely classical. He was afraid that quantum theory could invalidate his five-dimensional approach, as he explicitly stated at the end of his paper (\cite{Kaluza-trad}; p. 8).}. The most well-known contribution was Oskar Klein's\footnote{The modern multidimensional approach used by e.g. supergravity and string theory is called Kaluza-Klein approach in honour of these two authors, but the modern approach is different from that of the Fathers. For a review of the modern approach and a comparison with the old one see \cite{Duff-86}.} work, who developed his ideas from 1926 to 1927. Less known contributions were the papers written by Louis de Broglie \cite{deBroglie} and L\'eon Rosenfeld \cite{Ros1} \cite{Ros2} \cite{Ros3} \cite{Ros4} \cite{Ros5}. During the year spent in Paris, Rosenfeld started to interact frequently with de Broglie, discussing for example the problem of spin. It was the Belgian physicist who drew de Broglie's attention on the five-dimensional approach. As a consequence the French physicist published a paper, in 1927, on this topic \cite{deBroglie} \cite{Kuhn1}. During Kuhn's interview Rosenfeld also recollected that he was anxious to apply his new acquired knowledge to relativity, and that the first goal he wanted to achieve was to develop `the wave equation in five dimensions' \cite{Kuhn1}. On this subject Rosenfeld published two notes during his stay at the Ecole Normal in Paris: \cite{Ros1} and \cite{Ros2}. Why did Rosenfeld decide to embark on a five-dimensional adventure? What attracted him? What was Rosenfeld's point of view at that time? In the case of Klein's work the answer was well known, because the Swedish physicist himself answered the question. As we will see, Klein, de Broglie and Rosenfeld constructed their five-dimensional approaches starting from different perspectives and we will try to make clear what considerations suggested to each of the three authors how to develop a five-dimensional picture.

Another important role for the young Rosenfeld was played by Th\'eophile De Donder. Like Rosenfeld, De Donder was a Belgian researcher, older and more experienced. De Donder was an enthusiastic supporter of Einstein's theory. As we will see, soon after the birth of QM he tried to explain the existence of atomic stable orbits with the help of GR, but he always followed a classical approach \cite{DD1} \cite{DD2} \cite{DD3}. As Rosenfeld recollected: `I published a note which I sent to him to be presented to the Belgian Academy. De Donder was the least critical person you can imagine, he was enthusiastic about it. So he asked me then to come to Brussels, he wanted to have me in Brussels; I wanted to go abroad a bit more, but I worked for a month with him in Brussels.' \cite{Kuhn1}. As we shall see, one of the main consequences of the Rosenfeld-De Donder collaboration in 1927 was the physical interpretation of the assumptions made by Rosenfeld in his first paper, with the introduction of Bohr's correspondence principle in the context of QG, contained in \cite{Ros3} \cite{Ros5} \cite{DD-Ros}. In October 1927 the fifth Solvay conference took place in Brussels and on that occasion De Donder tried to attract attention to Rosenfeld's work. This Solvay conference is well known to historians of Physics, because it indicates the start of the famous Einstein-Bohr debate. The young Belgian physicist was not officially admitted to attend the conference, but de Donder invited Rosenfeld to follow him. At the conference Rosenfeld met Max Born for the first time and asked him about the possibility of a stay in G\"ottingen. Born's positive answer permitted Rosenfeld to attend Hilbert's, Born's and Pascual Jordan's lectures (\cite{Rosenfeld-bio}, p. 18), and it would open the doors to his future collaborations with Pauli, Jordan and many others. All these facts showed the crucial role played by De Donder in Rosenfeld's life.

In the next section we will start with a brief summary of the history of Klein's work and its intersection with de Broglie's contribution to the construction of a five-dimensional Universe. Section 3 will end with an introduction of De Donder four-dimensional approach, based on the lectures he gave at MIT in 1925, in order to understand, in section 4, how De Donder also influenced Rosenfeld's early work.

\section{Oskar Klein's, Louis de Broglie's and Theophile De Donder's role}\label{5D}
\subsection{The five-dimensional Universe: Klein's approach}\label{section-Klein}
Klein's investigation of the five-dimensional Universe started in 1926 with the purpose of unifying gravity, electromagnetism and WM \cite{Pais2}. 
As Klein himself recollected in \cite{Klein-life}, he was attracted by two facts. First, he knew that the Hamilton-Jacobi (HJ) equation offers a link between particle dynamics and the propagation of a wave front, in the limit of geometrical optics, suggesting a concrete realization for the wave-particle duality. Secondly, by writing the relativistic HJ equation for a particle moving in a combined gravitational and electromagnetic field, he noticed that the electric charge would play the role of an extra momentum component: `[...] I gave a lecture course on electromagnetism, towards the end of which I derived the general relativistic Hamilton-Jacobi equation for an electric particle moving in a combined gravitational and electromagnetic field. Thereby, \textit{the similarity struck me between the ways the electromagnetic potentials and the Einstein gravitational potentials enter into this equation}, the electric charge in appropriate units - appearing as the analogue to a fourth momentum component, the whole looking like a wave front equation in a space of four dimensions. [emphasis added]'\footnote{It is worth noting that in the original paper Klein did not emphasize the role of the electric charge explicitly. Rosenfeld followed a similar reasoning in constructing his wave equation, but stated it explicitly: see the remark after equation (\ref{eq-cl2}).} (\cite{Klein-life}; p. 108)\footnote{The original reasoning runs backward with respect to the path followed by Klein in the paper, where the author presented his model in an axiomatic way.}. In the summer of 1925 he became `immediately very eager to see how far the mentioned analogy reached' (\cite{Klein-life}; p. 109) and he started to investigate the five-dimensional Riemann geometry to describe the gravitational and electromagnetic interactions in a unified framework, trying also to write a five-dimensional wave equation. In the long wavelength limit, the wave equation resembles the eikonal equation for the paths of light rays in geometric optics. These paths follow geodesic lines through a Riemannian space: Klein identified them with five-dimensional null-geodesics which reduce, on his assumptions, to four-dimensional trajectories for charged massive particles moving in a combined electromagnetic and gravitational field. Klein's original idea was to follow an analogy with light in five dimensions, even if he wanted to relate five-dimensional geometry with the stationary states of massive particles. Carrying on this work, the Swedish Physicist convinced himself that his approach was only a first step towards the formulation of a theory able to reconcile GR with WM. But this conclusion was contained only in his last paper of the period \cite{Klein5}, a work that Rosenfeld would never cite.

Now we briefly retrace the steps followed by Klein in his first paper \cite{Klein1} \cite{Klein1-trad}. Klein introduced the following five-dimensional line element\footnote{In our paper we consider many authors who introduced different notations. We decided to adopt the following conventions. Barred indices refer to the five-dimensional World, $ \bar\mu = 0, 1, 2, 3, 5$, where the zero component corresponds to a time-like dimension. We use the mostly-plus signature, i.e. $\eta_{\bar\mu\bar\nu}=diag(-1,+1,+1,+1,+1) $. The unbarred Greek indices correspond to the usual four-dimensional space-time, $ \mu = 0,1,2,3 $, and Latin indices refer to the three-dimensional spatial coordinates, $ i = 1,2,3 $. We use International System of Units.}:
\begin{equation}\label{d-sigma} 
d\sigma^2 = \gamma_{\bar{\mu}\bar{\nu}}dx^{\bar{\mu}}dx^{\bar{\nu}}\, ,
\end{equation}
assuming that the metric tensor did not depend on the new fifth space-like component\footnote{Kaluza called this hypothesis the \textit{cylinder condition}. Using modern language, this means that translations in the fifth direction are isometries and hence that the five-dimensional space-time admits a space-like Killing vector field, namely $ \frac{\partial}{\partial x^5} $. Neither Klein nor de Broglie or Rosenfeld mentioned this fact explicitly in their papers.\label{cylinder}} $ x^5 $. Then it follows that the allowed coordinate transformations were restricted to the following set:
\begin{subequations}  
	\begin{empheq}
	[left={\empheqlbrace} ]{align}
	x^{\mu} &= f^{\mu}\left(  {x^0}^{\prime}, \, {x^1}^{\prime}, \, {x^2}^{\prime},\, {x^3}^{\prime}\right) \label{transform1}\\
	x^5 &= x^{5\prime} + f_5\left(  {x^0}^{\prime},\, {x^1}^{\prime}, \, {x^2}^{\prime},\,  {x^3}^{\prime}\right)  \, .\label{transform2}
	\end{empheq}
\end{subequations}
(\cite{Klein1-trad}; p. 11). After noting the invariance of $  \gamma_{55} $ under the coordinate transformations (\ref{transform1}) and (\ref{transform2}), Klein decided to set  $ \gamma_{55}=\alpha $, where $ \alpha $ is a constant. In modern Kaluza-Klein theories $  \gamma_{55} $ is not a constant, it is a real scalar field depending on the transverse dimensions, called a dilaton field. As Lochlain O'Raifeartaigh \cite{Raife2000} and other authors \cite{KK-rev} pointed out, Klein's choice is inconsistent, as we shall explain below after equations (\ref{eq-moto-penta}). 
Klein rewrote the line element (\ref{d-sigma}) in the following form:
\begin{equation}\label{nota-d-sigma1}
d\sigma ^2  = \alpha d\theta ^2 + ds^2 \, , 
\end{equation}
\begin{center}
where
\end{center}
\begin{equation}\label{nota-d-sigma2}
d\theta = dx^5 + \frac{\gamma_{5\mu}}{\alpha}dx^{\mu}\quad ;\quad
g_{\mu\nu} = \gamma_{\mu\nu} - \frac{\gamma_{5\mu}\gamma_{5\nu}}{\alpha}\quad ;\quad ds^2=g_{\mu\nu}dx^{\mu}dx^{\nu}\; \, .  
\end{equation}
Citing Kramers' paper on stationary gravitational fields in four dimensions \cite{Kramers}, Klein noted that $ d\theta$, equation (\ref{nota-d-sigma2}), is invariant under the coordinate transformations (\ref{transform1}) and (\ref{transform2}). In fact, following Kramers and remembering that $ \alpha=\gamma_{55} $, the invariance of $ d\theta $ is transparent if we rewrite it in the following way: $\displaystyle d\theta = dx^5 + \frac{\gamma_{5\mu}}{\gamma_{55}}dx^{\mu} = \frac{1}{\gamma_{55}}\gamma_{5\bar{\mu}}dx^{\bar{\mu}} $. As a consequence, Klein noted that the four components $ \gamma_{5\mu} $ transform as a four-vector of the four-dimensional space-time. Following Kaluza, Klein assumed that they would be proportional to the electromagnetic potentials $ A^\nu = (V;\vec{A}) $, introducing another parameter $ \beta $:
\begin{equation}\label{def-beta}
\frac{\gamma_{5\mu}}{\alpha} = \beta A_{\mu} \; ,
\end{equation}
where we defined $ A_\mu = g_{\mu\nu}A^\nu $. We note that $ d\theta $ defined in equation (\ref{nota-d-sigma2}) is not an exact form and that it can be rewritten as: $ d\theta = dx^5 + \beta A_{\mu}dx^{\mu} $. Using $ d\theta^2 $ invariance and $ d\sigma^2 $ invariance, it follows that $ ds^2 $ is invariant under the coordinate transformations (\ref{transform1}) and (\ref{transform2}). As a consequence $ g_{\mu\nu} $ can be interpreted as a four-dimensional metric. After having introduced the five-dimensional curvature scalar$ \tilde{R} $, defined in appendix \ref{app1}, Klein varied the five-dimensional Einstein-Hilbert action as usual in GR, with respect to the metric $ \gamma_{\bar{\mu}\bar{\nu}} $:
\begin{equation}\label{variation}
 \delta_{\gamma}\mathcal{S}_5 = \delta_{\gamma} \int_{\Omega} \tilde{R}\sqrt{-\gamma} d^5x  =  \int_{\Omega} d^5x \frac{\delta\left( \tilde{R}\sqrt{-\gamma}\right) }{\delta \gamma_{\bar{\mu}\bar{\nu}}}\delta \gamma_{\bar{\mu}\bar{\nu}}\; ,
\end{equation} 
where the symbol $ \sqrt{-\gamma} $ represents the square root of the negative of the determinant of the metric and the integral is carried out over a closed region $ \Omega $, where boundary values of $ \gamma_{\bar{\mu}\bar{\nu}} $ are kept fixed. From the principle of stationary action the five-dimensional Einstein equations follow:
\begin{equation}\label{Klein-5-action}
\delta_{\gamma}\mathcal{S}_5 = 0 \quad\Rightarrow\quad \tilde{R}_{\bar\mu\bar\nu}-\frac{1}{2}\gamma_{\bar\mu\bar\nu}\tilde{R}=0\;\, .
\end{equation}
It is worth noting that neither Klein nor any of the other authors we analysed considered the $ 55 $ component of equation (\ref{Klein-5-action}), because they fixed $ \alpha = \text{constant} $ before varying the action. Thanks to all assumptions he made, equations (\ref{Klein-5-action}) are formally equivalent to the four-dimensional Einstein-like equations coupled to the four-dimensional Maxwell-like equations\footnote{See appendix \ref{app3c} for a detailed explanation of the formal equivalence in the context of Rosenfeld's work.}:
\begin{subequations}\label{eq-moto-penta}  
\begin{empheq}
[left={\empheqlbrace} ]{align}
{R}_{\mu\nu}-\frac{1}{2}g_{\mu\nu}R &= \frac{\alpha\beta^{2}}{2} T_{\mu\nu}^{em} \label{eq-kla} \\
\partial_{\mu}\left( \sqrt{-g}F^{\mu\nu} \right) &= 0  \label{eq-klb} \quad ,
\end{empheq}
\end{subequations}
where $ g $ is the determinant of $ g_{\mu\nu} $ defined in equation (\ref{nota-d-sigma2}). Choosing to set\footnote{In his following papers Klein would set $ \alpha = 1 $. In de Broglie's and Rosenfeld's paper both constants are present.} $ \alpha\beta^{2}=\frac{16\pi G}{c^4} $, where $ G $, and $ c $ are the Newton constant and the speed of light respectively, Klein justified the identification of $ g_{\mu\nu} $ and  $ F_{\mu\nu}=\partial_\mu A_\nu - \partial_\nu A_\mu $ with our four-dimensional metric and with the electromagnetic tensor respectively. The electromagnetic stress-energy tensor that appears in (\ref{eq-kla}) is defined by: $ T_{\mu\nu}^{em}={F_\mu}^\alpha F_{\nu\alpha}-\frac{1}{4}g^{\mu\nu}F_{\alpha\beta}F^{\alpha\beta} $. The condition $ \alpha\beta^{2}=\frac{16\pi G}{c^4} $ implies $ \alpha > 0 $. This means that Klein introduced a space-like extra dimension motivated by the need to obtain the four-dimensional Einstein equations coupled with Maxwell's equations. Indeed, a space-like coordinate only, i.e. a positive $ \alpha $ constant in (\ref{eq-kla}), produces the correct coupling between electromagnetic and gravitational interactions.
 In this sense our four-dimensional World is a ``projection'' of a five-dimensional Universe.

As indicated Klein's model is inconsistent, if $ \alpha $ is constant. Indeed, if the dilaton is a non trivial scalar function $ \alpha (x) $, the $ 55 $ component of equations (\ref{Klein-5-action}) is not trivial and it has the form $ \displaystyle \square \sqrt{\alpha}\sim \left( \sqrt{\alpha}\right)^3 F_{\alpha\beta}F^{\alpha\beta} $, where the four-dimensional operator $ \square $, when acting on a scalar function $ \alpha(x) $ is defined by $ \square \alpha =g^{\mu\nu}\nabla_{\mu}\partial_{\nu}\alpha $ for a curved four-dimensional space-time, where $ \nabla_{\mu} $ represents the covariant derivative. This means that a non-zero constant dilaton would imply the too restrictive condition $ F_{\alpha\beta}F^{\alpha\beta} = 0 $, i.e. that the modulus of the electric field should be proportional to the modulus of the magnetic field. As reported in \cite{KK-rev}, this inconsistency was noted by Pascual Jordan \cite{J1} and Yves Thiry \cite{T1} in 1947 and in 1948 respectively: all the authors of the period we are considering imposed the constancy of the dilaton, including de Broglie and Rosenfeld, and they were not aware of this inconsistency.

In order to reconcile this framework with WM, Klein's idea was to write a five-dimensional wave equation in a curved space-time, which was then to be connected with the classical four-dimensional Lorentz equation for a charged particle in the presence of gravitational and electromagnetic fields, in the so called geometrical optics limit. The connection between the two equations, considered by all the authors that we shall analyse, is as follows\footnote{For a short review with some mathematical details see appendix \ref{geom-optic}. For a detailed technical explanation of Klein's approach see e.g. \cite{Raife2000}.}. In a geometrical optics approximation, the wave equation reduces to the classical HJ equation with a particular Hamiltonian function. After a Legendre transformation, the associated Lagrangian produces five equations of motion. The four equations transverse to the fifth coordinate can be reduced to the Lorentz equation for a charged massive particle. The Lagrangian approach shows that, in five dimensions, charged particles follow a geodesic motion. Klein himself explained this procedure in the introduction of his paper: `the equations of motion for the charged particles [..] take the form of equations of geodesic lines. If we explain these equations as wave equations because the matter is supposed to be a kind of wave propagation, we are almost naturally led to a partial differential equation of second order, which may be regarded as a generalization of the ordinary wave equation.' (\cite{Klein1-trad}; p. 10). This justifies Klein's idea stated above to connect wave equation with geodesic lines and it also clarifies why WM had a prominent role in his approach in unifying GR with QM. 

In order to write an equation that generalizes Schr\"odinger's equation, Klein followed an analogy with light. The equation he found resembles a massless Klein-Gordon (KG) equation\footnote{Given a scalar field $ \phi $ of mass $ m $, the KG equation is $ \square\phi = \frac{m^2 c^2}{\hbar^2}\phi  $.}, what the author called `our equations for the light wave' (\cite{Klein1-trad}; p. 17). The Swedish physicist was forced to introduce a symmetric tensor $ a_{\bar\mu\bar\nu} $, whose contravariant components are fixed by the request to connect the five-dimensional wave equation with the four-dimensional Lorentz equation for massive charged particles, as we shall see below. Klein's wave equation reads:
\begin{equation}\label{KG1} 
a^{\bar{\mu}\bar{\nu}}\left( \delta^{\bar\sigma}_{\bar\nu}\frac{\partial}{\partial x^{\bar{\mu}}} -\Gamma_{\bar{\mu}\bar{\nu}}^{\bar{\sigma}}\right)\partial_{\bar{\sigma}}\Psi= a^{\bar{\mu}\bar{\nu}}\nabla_{\bar{\mu}}\partial_{\bar{\nu}}\Psi=0\; ,
\end{equation}
where he introduced the covariant derivative $ \nabla_{\bar{\mu}} $ using the Christoffel symbols $  \Gamma_{\bar{\mu}\bar{\nu}}^{\bar{\sigma}}$, because Klein considered a wave function living on a curved five-dimensional Riemannian manifold. This means that Klein's wave function is different from Schr\"odinger's wave function, which lives in configuration space. With this respect, Klein's $ \Psi $ resembles a classical scalar field. From a modern point of view, the introduction of $ a^{\bar\mu\bar\nu} $ sounds strange, because the covariant derivative is usually contracted with the contravariant components of the metric $ \gamma^{\bar\mu\bar\nu} $, which are
different from $ a^{\bar\mu\bar\nu} $, as we shall see below. It is worth noting that Klein did not start from a variational principle to obtain his wave equation. He simply wrote a light-like wave equation. The hypothesis that the wave function would be periodic with respect to the fifth coordinate $ x^5 $ permits to ``project'' equation (\ref{KG1}) to obtain the KG wave equation\footnote{Klein and all the authors we consider in the present paper were convinced, at that time, that the relativistic wave equation for the electron would be the KG equation, instead of Dirac's equation. It is worth remembering that Pauli matrices were introduced in the same year \cite{Pauli} and that the Dirac's equation would be published one year later \cite{Dirac}.}. See appendix \ref{app2} for an explanation of the use of periodicity condition in the context of de Broglie's work. 

How did Klein justify the analogy with light? In \cite{Klein-life} the author recollected: `[...] for some time I had played with the idea that \textit{waves representing the motion of a free particle had to be propagated with constant velocity, in analogy with light waves} - but in a space of four dimensions - so that the motion we observe is a projection on our ordinary three-dimensional space of what is really taking place in four-dimensional space. [emphasis added]' (\cite{Klein-life}; p. 108).
The introduction of the symmetric tensor $ a^{\bar{\mu}\bar{\nu}} $ served this specific purpose. Klein's conviction was enforced by the fact that in the long wavelength limit equation (\ref{KG1}) reduces to the eikonal equation for light rays. As a consequence, Klein imposed that in the semi-classical limit the four-dimensional motion of charged particles with mass $ m $ in the presence of a gravitational and electromagnetic field should be described by five-dimensional null-geodesics of the following differential form: 
\begin{equation}\label{def-a}
d\hat{\sigma}^2 = a_{\bar{\mu}\bar{\nu}}dx^{\bar{\mu}}dx^{\bar{\nu}} = \frac{1}{m^2c^2}d\theta^2 + ds^2	
\end{equation}
(\cite{Klein1-trad}; p. 17) and showed that the correspondent geodesic equation is equivalent to the four-dimensional Lorentz equation. It seems that Klein introduced a different metric for the microscopic world, $ a_{\bar\mu\bar\nu} $, whose components can be obtained from equation (\ref{def-a}), namely:
\begin{equation}\label{metric-K} 
a_{\mu\nu} = g_{\mu\nu}+\frac{e^2}{m^2c^4}A_\mu A_\nu \quad\quad a_{\mu 5}=\frac{e^{2}}{m^2c^3\beta}A_\mu \quad\quad a_{55}= \frac{e^2}{m^2c^4\beta^2}  \;\, ,
\end{equation}
and which is quite unlike the space-time metric $ \gamma_{\bar{\mu}\bar{\nu}} $, cfr. eq. (\ref{metric-K}) with (\ref{nota-d-sigma2}) and (\ref{nota-d-sigma1}), but he made no comments on this choice. It is worth noting that the particle's mass $ m $ and its charge $ e $ are hidden in the expressions of $ a_{\bar{\mu}\bar{\nu}} $ tensor.

To show the correspondence between five-dimensional null-geodesics and four-dimensional motion of charged particles, Klein considered the corresponding Lagrangian picture, by projecting the equations of motions obtained by varying the Lagrangian $\displaystyle  L = \frac{1}{2} a_{\bar\mu\bar\nu}\frac{dx^{\bar\mu}}{d\hat{\lambda}}\frac{dx^{\bar\nu}}{d\hat{\lambda}} $, where $ \hat{\lambda} $ is an arbitrary parameter. One of the five resulting Euler-Lagrange equations states that the momentum conjugated to the coordinate $ x^5 $ is conserved, while the other four equations are equivalent to the Lorentz equation for an electron\footnote{Technical details of the equivalence are given in appendix \ref{geom-optic}.} (charge $ q=-e $):
\begin{equation}\label{eq-lorentz} 
mc\left( \frac{d}{d\tau}\left( g_{\mu\nu}u^{\nu}\right) -\frac{1}{2}\partial_\mu g_{\rho\nu}u^\rho u^\nu\right) =-\frac{e}{c}\left( \partial_\mu A_\nu-\partial_{\nu}A_{\mu}\right) u^\nu\, ,
\end{equation} 
where the four-dimensional proper time $ \tau $ is defined by $ d\tau= \sqrt{-ds^2}$, and the four-velocity of the particle is defined by $ \displaystyle u^{\mu}=\frac{dx^\mu}{d\tau} $.
The analogy with light forced Klein to look for a correspondence between five-dimensional null-geodesics and four-dimensional paths: this conclusion would be criticized by de Broglie.

Before going on, it is worth noting that equation (\ref{eq-lorentz}) can be obtained, as Klein did, without fixing the constant\footnote{See appendix \ref{app2} for technical details in the context of de Broglie's work.} $ \beta $ introduced in (\ref{def-beta}). In his first paper, Klein decided to set $ \beta = \frac{e}{c}$ and consequently the value of $ \alpha $ must be $ \alpha=\frac{16\pi G}{e^2c^2} $.
In his second paper \cite{Klein2}, a brief communication to \textit{Nature}, it seems that Klein had changed his mind about the role of null-geodesics. In fact he explicitly referred to `the equation of geodetics' (\cite{Klein2}; p. 516) of the line element\footnote{In this brief communication Klein introduced a different notation and decided to set $ \alpha = 1 $ from the beginning and consequently $ \beta = \sqrt{\frac{16\pi G}{c^4}} $: this simply means that now the fifth coordinate has a dimension of length. \label{convention2}} $ d\sigma^2 $. Furthermore, he suggested to start from the new Lagrangian $\displaystyle  L^{\prime} = \frac{m}{2} \gamma_{\bar\mu\bar\nu}\frac{dx^{\bar\mu}}{d\tau}\frac{dx^{\bar\nu}}{d\tau} $, where the $ a_{\bar\mu\bar\nu} $ tensor has disappeared, and the mass and the presence of the proper time $ \tau $ indicate that Klein did not refer to null-geodesics\footnote{From a modern point of view, even in the massive case, the Lagrangian $ L^{\prime} $ should be written by introducing the arbitrary parameter $ \hat{\lambda} $. The proper time $ \tau $ can be introduced because the ratio $ \frac{d\hat{\lambda} }{d\tau}$ is constant, as we shall show in appendix \ref{app2}, discussing de Broglie's approach. We suppose that Klein underlined implicitly that he did not consider null-paths any more.}. This brief communication is important, because Klein noted that the quantization of the momentum along the periodic fifth dimension\footnote{The momentum connected with the quantization of the electric charge is $ p_5 $, the momentum conjugated to the fifth dimension, namely $ \displaystyle p_5 = \frac{\partial L^{\prime}}{\partial \left( dx^5 /d\tau\right)}\; $.\label{nota-hat-lambda}} of finite size $ l $ could have been connected with the quantization of the electric charge. In fact the momentum's quantization along the fifth dimension forces the size $ l $ to assume a precise value: 
\begin{equation}\label{Klein-l}
l=\frac{hc\sqrt{2\kappa}}{e},
\end{equation}
where $\displaystyle \kappa = \frac{8\pi G}{c^{4}}$. As we will see, as far as we know, neither de Broglie nor Rosenfeld fixed explicitly either of both parameters and they also did not make explicit considerations on the size of the fifth dimension.

\subsection{De Broglie's contribution}\label{section-deBroglie}
As mentioned in the introduction, during his stay in Paris Rosenfeld drew de Broglie's attention to Klein's approach. From de Broglie's point of view, the analogy with light was not the correct perspective to describe the path of massive particles. In order to explain the conclusion reached by de Broglie, we emphasize again that Klein, de Broglie and Rosenfeld developed the five-dimensional Universe for different reasons. 

De Broglie's paper analyses the features of the five-dimensional approach from two distinct perspectives: the classical and the quantum point of view. In the first part of de Broglie's paper, the author described how the most attractive advantage of the classical five-dimensional approach would reside in the fact that it allowed to geometrize all the forces known at that time, i.e. the gravitational and the electromagnetic forces. The author made an analogy between Einstein's approach and the five-dimensional construction. De Broglie interpreted Einstein's theory as a geometrical description of the gravitational force and Kaluza's approach as an extension of this geometrical description to Maxwell's theory\footnote{Here and in the following, we present an English translation of some parts of the original paper, written in French.}: `The main consequence of the introduction of the equivalence principle is that the metaphysic notion of force in the theory of gravitation disappears. The path followed by a point particle in a gravitational field can be defined, thanks to Einstein's conceptions, as the geodesic line of the space-time. [...] The success of this beautiful interpretation of the gravitational field temptingly suggests to throw out the concept of force from the Physics, in order to replace it with the concept of geometry.' (\cite{deBroglie}; p. 65). 

In the second part of the paper, de Broglie introduced the description of the quantum behaviour of matter using wave/particle duality. From this perspective, there are no forces associated to the particles' wave function, hence neither geometrical description nor analogy with light was needed. De Broglie explicitly stated that `With the present state of our knowledge it seems that all the forces of which we are aware can be reduced to only two: the gravitational and electromagnetic forces.' (\cite{deBroglie} p. 65). It is worth noting that the quantum force concept emerged with the introduction of quantum fields. Unlike Klein, de Broglie introduced a wave equation describing quantum particles' dynamics, i.e. the KG equation, in four dimensions: in the geometrical optics approximation the wave's rays would follow the classical trajectories for massive particles. Hence a five-dimensional generalization of the KG equation would not require any analogy with light. It is important to stress that de Broglie did not use any variational principle to describe the wave's dynamics. With this premise in mind we first consider de Broglie's approach in more detail. 

De Broglie briefly reviewed Klein's approach and introduced the line element (\ref{d-sigma}) with Klein's Ansatz that now we rewrite here for convenience:
\begin{eqnarray}\label{d-sigma2}
d\sigma ^2  &=& \alpha d\theta ^2 + ds^2 \, , \label{nota-d-sigma1b}\\
&\text{where}&\nonumber\\
d\theta = dx^5 + \beta A_{\mu}dx^{\mu}\quad ;\quad
g_{\mu\nu} &=& \gamma_{\mu\nu} - \frac{\gamma_{5\mu}\gamma_{5\nu}}{\alpha}\quad ;\quad ds^2=g_{\mu\nu}dx^{\mu}dx^{\nu}\; \,   \label{nota-d-sigma2b}
\end{eqnarray}
(We adapted de Broglie's notation changing the symbols he used). Let the values of $ \alpha $ and $ \beta $ be unfixed for the moment. De Broglie's choice shall be analysed after equation (\ref{Iquadro}).

At this point, de Broglie's and Klein's paths separate. As we said, de Broglie did not consider any analogy with light, hence he studied the geodesic equations in five dimensions for massive particles. Like Klein, the key idea is that our world would be a projection onto a four-dimensional manifold of what happens in the five-dimensional Universe. The four-dimensional geodesic equation is obtained by the following variational principle\footnote{Because of our mostly-plus signature, the four-dimensional action for a point particle involves the proper time $ \tau $.}:
\begin{equation}\label{azione4}
\delta S_4=0\quad\Rightarrow\quad\delta\, \int_{O}^{M} d\tau = 0\, ,
\end{equation}
where $ O $ and $ M $ are `two fixed points of the world line' (\cite{deBroglie}; p. 69).
De Broglie considered its natural generalization in five dimensions:
\begin{equation}\label{azione5}
\delta S_5=0\quad\Rightarrow\quad\delta\int_{O}^{M} d\hat{\tau} = 0\, ,
\end{equation}
where we introduced the notation $ d\hat\tau = \sqrt{-d\sigma^{2}} $.
The geodesic equations following from (\ref{azione5}) are equivalent to the five-dimensional equations obtained by Klein with the help of the $ a_{\bar{\mu}\bar{\nu}} $ tensor he introduced in his first paper\footnote{See appendix \ref{app2} for a detailed explanation of the original derivation. As we said, Klein was certainly aware of this fact, because he changed his own approach to the geodesics in the brief communication to \textit{Nature}. It is worth noting that de Broglie never cited Klein's \textit{Nature} paper.}, and their four-dimensional projection reproduce equations (\ref{eq-lorentz}). In order to obtain the correct Lorentz equations, de Broglie set
\begin{equation}\label{ratio}
	\alpha\frac{d\theta}{d\tau} = -\frac{e}{\beta c}\frac{1}{mc}\; ,
\end{equation}
underlining the importance of this equation. Indeed, from de Broglie's point of view, equation (\ref{ratio}) suggests a geometrical interpretation of the ratio $ \frac{e}{m} $. Let's consider, following de Broglie, `a coordinate line $ x^5 $' (\cite{deBroglie}; p. 68) and using $ d\tau =\sqrt{-ds^2} $ and $ d\hat{\tau} = \sqrt{-d\sigma^2} $ we rewrite equation (\ref{nota-d-sigma1b}) as follows:
\begin{equation}\label{d-sigma3}
d\hat{\tau}^2 = d\tau^2 +\left| \alpha\right|  d\theta^2\; .
\end{equation}
We use $\left| \alpha \right| $, because de Broglie set $ \alpha <0 $, a choice that we shall discuss after equation (\ref{Iquadro}). `Let us represent, on a point $ P $ of this coordinate line, a part of a plane $ \pi $ inclined with respect to the $ x^5 $ direction, which represents a little portion of the four-dimensional hypersurface $ x^5 = const. $ passing through the point $ P $. Let $ \overline{PQ} $ be an element of a world line of length $d\hat{\tau} $ and let $ \overline{PS} $ and $ \overline{PR} $ be its  projections along the $ x^{5} $ direction and orthogonal to the $ x^5 $ direction respectively. From equation (\ref{d-sigma3}) it follows that
\begin{equation}\label{projection}
\overline{PS} = \sqrt{\left| \alpha\right| } d\theta	\;\text{;}\qquad\overline{PR} = d\tau\; .
\end{equation}
[...] the tangent of the angle $ \widehat{QPR}  $, namely $\displaystyle
 \frac{\sqrt{\left| \alpha\right| } d\theta}{d\tau} $, is proportional to the ratio $ \frac{e}{m} $ where $ e $ and $ m $ are the charge and the mass of the particle of which $  \overline{PQ} $ is the element of the world line. Hence the world line of every moving object makes the same angle with the direction $ x^5 $ at each point, which angle is straight if the electric charge is zero.' (\cite{deBroglie}; p. 68)\footnote{With the choice $ \alpha > 0 $, the ratio would define the hyperbolic tangent of the angle.}. This result supported de Broglie's conviction that the five-dimensional Universe could provide a geometrical description for all of the known physical concepts. Rosenfeld would continue to use this idea, as we shall see in the discussion after equation (\ref{slope}).

De Broglie asked himself what the exact form of the action $ S_5 $ to be varied would be in order to obtain a five-dimensional generalization of the four-dimensional massive particle's action. 
De Broglie stressed that he wanted to obtain, in the case of zero charge, the usual action $\displaystyle S_4 = -mc\int_{O}^{M} d\tau $ (\cite{deBroglie}; p. 70) and he proposed that the five-dimensional particle's action should be\footnote{We skip over some technical details. See the appendix \ref{app2} for de Broglie's original proof that $ S_5 $ reduces to $ S_4 $ in the case of null charge.}:
\begin{equation}\label{action-dB}
S_5 =-\mathcal{I} \int_{O}^{M} d\hat\tau\, ,
\end{equation}
where the quantity $ \mathcal{I} $ 
satisfies the following relations
\begin{equation}\label{relazioni}
\mathcal{I}\alpha \frac{d\theta}{d\hat\tau} = -\frac{e}{c\beta}\, , \quad\quad 
\mathcal{I}\frac{d\tau}{d\hat\tau} = mc
\, ,
\end{equation}
and has the following form: 
\begin{equation}\label{Iquadro}
\mathcal{I}=\sqrt{m^2c^2 - \frac{e^2}{\alpha\beta^{2} c^{2}} }\, .
\end{equation} 
  
The invariant $ \mathcal{I} $ needs some comments, connected with de Broglie's choice of $ \alpha $'s and $ \beta $'s values. De Broglie implicitly set
\begin{equation}\label{alfa-dB}
\alpha\beta^{2}=-\frac{16\pi G}{c^{4}}\; ,
\end{equation}
from the beginning of his paper. As a consequence, $\displaystyle \mathcal{I}_{dB} = \mathcal{I}\left( \alpha\beta^{2}=-\frac{16\pi G}{c^{4}}\right)$ is a real constant:
\begin{equation}\label{Iquadro-dB}
\mathcal{I}_{dB}=\sqrt{m^2c^2 + \frac{e^2c^2}{16\pi G} }\, ,
\end{equation}
and comparing $ S_4 $ and $ S_5 $, de Broglie suggested that it should be interpreted as the modulus of the five-dimensional momentum $ P_{\bar{\mu}} $ for charged particles, defined in analogy with the four-dimensional momentum $\displaystyle p_{\mu}=mcg_{\mu\nu}\frac{dx^\nu}{d\tau} $ for uncharged particles in four dimensions, namely $\displaystyle P_{\bar{\mu}} = \gamma_{\bar{\mu}\bar{\nu}}\mathcal{I}_{dB}\frac{dx^{\bar{\nu}}}{d\hat{\tau}}$ . To be more explicit, referring to the geometrical picture discussed above, de Broglie asserted that relations (\ref{relazioni}) should be interpreted as the tangent and orthogonal components of the five-dimensional momentum $ P_{\bar{\mu}} $ with respect to the fifth direction $ x^5 $ (\cite{deBroglie}; p. 70, note $ (^1) $). We will return to this interpretation discussing Rosenfeld work, see the discussion after equation (\ref{slope}). Equation (\ref{alfa-dB}) means that unlike Klein, de Broglie imposed that the fifth dimension would be a time-like coordinate, because from equation (\ref{alfa-dB}) it follows $ \gamma_{55}=\alpha <0 $. De Broglie made no explicit comment on the time-like character of the fifth dimension. As we shall see, Klein noted that this choice was inconsistent with other demands of the model. Rosenfeld would be strongly influenced by de Broglie's ideas, but he was aware of this inconsistency. After having specified this fundamental difference between the two approaches, let us now return to de Broglie's considerations.

After having established that the Lorentz equations (\ref{eq-lorentz}) can be obtained by varying\footnote{See appendix \ref{app2}.} $ \mathcal{S}_{5} $, de Broglie declared: `\textit{The notion of force has been banned completely from Mechanics.}' (\cite{deBroglie}; p.70), emphasizing his original aim. As a consequence he proposed the following wave equation as a generalization of Schr\"odinger wave equation, instead of (\ref{KG1}), namely
\begin{equation}\label{KG4} 
\gamma^{\bar{\mu}\bar{\nu}}\nabla_{\bar{\mu}}\partial_{\bar{\nu}}\Psi\,=\,\frac{4\pi^2}{h^2}\mathcal{I}_{dB}^{2} \Psi \; ,
\end{equation} 
where now the covariant derivative is correctly contracted with the metric. Equation (\ref{KG4}) could resemble a KG equation in five dimension, where $ \frac{\mathcal{I}_{dB}}{c} $ plays the role of the mass in five dimensions, because it is a real quantity. It is worth noting that the identification of $ \Psi $ as a wave function prevents the identification of $ \mathcal{I}_{dB} $ with a mass term in the sense of modern field theory. 
Using the fact that the action $ S_5 $ can be rewritten as follows
\begin{equation}\label{action-dB1}
S_5=-\int_O^M \frac{e}{c\beta}dx^5+ \frac{e}{c}\int_O^M A_\mu dx^{\mu} -mc\int_O^M d\tau \; ,
\end{equation}
de Broglie could show that equation (\ref{KG4}) is equivalent to the four-dimensional KG equation for massive particles, which reduces to Schr\"odinger equation in the non-relativistic limit. In order to demonstrate his claim, de Broglie introduced the geometrical optics approximation, writing the five-dimensional wave function $ \Psi $ as
\begin{equation}\label{deB-optic}
\Psi = Ce^{\frac{i}{\hbar}S_5} = f(x,y,z,t)e^{\frac{i}{\hbar}\frac{ex^5}{c\beta}}
\end{equation} 
(\cite{deBroglie}; p. 72), where $ C $ is a constant and $ S_5 $ is the five-dimensional action defined in (\ref{action-dB1}). It is worth noting that De Broglie considered $ S_5 $ as an Hamiltonian action. This means that he interpreted the five-dimensional action as a ``Jacobi function''. As we will see, De Donder will be more explicit on this fact. At this point, De Broglie expressed his opinion on the analogy with light introduced by the Swedish physicist: `O. Klein writes the equation (\ref{KG4}) without the second member, and he concludes that the world-lines must be null-geodesics; it is in our opinion that the second term of (\ref{KG4}) is fundamental and that the world-lines are still geodesics, but not null-geodesics' (\cite{deBroglie}, p. 72; we modified the number of the cited equation in order to fit with our numerical order).

Before going on we return to the question of the fifth dimension's size, which was never calculated by de Broglie. Indeed, the author commented on the size of the fifth dimension like this: `The variations of the fifth coordinate completely escape our senses [...] two points that differ only for the value of the fifth coordinate are indistinguishable from our point of view' (\cite{deBroglie}; p.67). But from these observations, de Broglie inferred, like Klein, that the components of the metric $ \gamma_{\bar{\mu}\bar{\nu}} $ must be independent from the fifth coordinate and that `the only humanly possible transformations have the following form:
\begin{equation}\label{tir}
{x'}^\mu = f^\mu\left( x^0,\, x^1,\, x^2,\, x^3 \right)\quad \text{'}
\end{equation}
(\cite{deBroglie}; p.67). If de Broglie would have chosen $ \alpha\beta^2=2\kappa $, i.e. a space-like dimension, he would have been able to read off the size of the compact dimension. Indeed, after noting that\footnote{Remember that de Broglie choose a negative value for $ \alpha $. We suppose that for this reason he never noted this fact.} $\tilde{x}^5 = \sqrt{\alpha}x^5 $ has dimensionality of $ [length]^1 $, the dependence on the fifth dimension in (\ref{deB-optic}) can be rewritten as
\begin{equation}
\frac{i}{\hbar}\frac{ex^5}{c\beta}=\frac{i}{\hbar}\frac{e}{c\sqrt{\alpha}\beta}\sqrt{\alpha}x^{5}=\frac{i}{\hbar}\frac{e}{c\sqrt{2\kappa}}\sqrt{\alpha}x^{5}=i\frac{\tilde{x}^5}{\tilde{l}}\, ,
\end{equation}
where $ \tilde{l}=\frac{\hbar c\sqrt{2\kappa}}{e} $ is Klein's length (\ref{Klein-l}) divided by $ 2\pi $, showing that Klein's length determines the periodicity.

De Broglie was very impressed by equation (\ref{KG4}) and he concluded his paper with the following remark: `For studying the problem of matter and of its atomic structure deeply, it would be necessary to perform a systematic analysis of the five-dimensional Universe's point of view that seemed to be more promising than Weyl's approach. If we understand how to interpret correctly the role played by the constants $ e $, $ m $, $ c $, $ \hbar $ and $ G $ in equation (\ref{KG4}), we will have finally grasped one of the most mysterious secret of Nature.' (\cite{deBroglie} p.73).

Klein's answer to the question of null-geodesics arrived immediately \cite{Klein-risp-deBroglie}. He noted that in equation (\ref{KG4}) de Broglie used the metric $ \gamma^{\bar{\mu}\bar{\nu}} $ instead of his ``artificial'' tensor $ a^{\bar{\mu}\bar{\nu}} $: inserting the components of $ a^{\bar{\mu}\bar{\nu}} $ in (\ref{KG4}), Klein showed that the equations (\ref{KG4}) and (\ref{KG1}) were equivalent. The fact is not surprising, because the particle's mass is hidden in the expression of the $ a^{\bar{\mu}\bar{\nu}} $ tensor\footnote{See discussion after equation (\ref{def-a}).}. Klein also noted that the condition on the parameters $ \alpha\beta^2=2\kappa $ was incompatible with the choice of a time-like fifth dimension\footnote{In appendix \ref{app1} we will analyse Klein's claims in more detail.}. But he concluded the brief communication with a positive comment on de Broglie's assertion: `...this \textit{error} has no influence on de Broglie's result [emphasis added]'\footnote{Klein assertion was referred to the fact that irrespective of the nature of the fifth coordinate, after having used the periodicity condition, the term with the Newton constant in (\ref{KG4}) disappears and it reduces to the KG equation. See appendix \ref{app2}, the discussion after equation (\ref{psi1}) for a detailed explanation.} (\cite{Klein-risp-deBroglie}; p. 243). It is worth noting that in his subsequent papers Klein would have stressed the need to introduce a space-like fifth dimension\footnote{See appendix \ref{app1} for technical details.} (\cite{Klein5}; p. 206, footnote *). Notwithstanding, after de Broglie's paper, Klein abandoned explicitly the analogy with light.

\subsection{De Donder's lectures on gravitation}
Neither Klein nor de Broglie tried to obtain their wave equation, in the works we analysed so far, using a unified variational principle. In fact they introduced only the particle's Lagrangian in order to describe the classical particle's dynamics\footnote{As we shall note in the next section, Klein's last paper would contain a five-dimensional variational principle to derive WM (\cite{Klein5}; p. 201), which is slightly different from Rosenfeld's variational principle.}. The Belgian physicist Th\'eophile De Donder was an early supporter of variational principles, developing the purely formal parts of the calculus of variations and analysing e.g. the effect of transformations of coordinates and parameters upon what he called ``invariants'' and upon other  expressions which occur in the theory of the variational calculus \cite{DD-var}. As we shall see, De Donder's ``invariants'' would correspond to our modern Lagrangian density. He tried also to derive WM from a unified variational principle. He did not consider multidimensional world, because he was satisfied to write a unified Lagrangian involving the gravitational field, the Maxwell field and a Lagrange function for the quantum particle. De Donder tried to present a coherent framework for relativistic Lagrangian dynamics in the context of curved spaces, and he was one of the first to note the role of the HJ equation as constraints in this context. In his first paper, Rosenfeld mainly followed De Donder's approach to introduce the wave function in the five-dimensional Universe, as we shall see later.  
During the Spring of 1926, De Donder gave a series of lectures at the MIT. In these lectures, which would be published the following year \cite{DeDonder2}, the Belgian physicist gathered together all the results he had just published in the \textit{Comptes Rendus} journal. The lectures contain all the original references, with an advantage: \textit{Comptes Rendus} publications were often brief communications, whereas the lectures gave a complete overview of De Donder's point of view. For this reason we will refer to his MIT lectures. We stress that this paragraph is a brief analysis of the ideas that influenced Rosenfeld. A deeper understanding of De Donder's methods goes beyond the goals of the present paper.  

The Belgian physicist tried explicitly to apply GR to the microscopic world. At the end of the first lecture, the general introduction, De Donder wrote: `We then say a few words about the mysterious quantum. To shed some light on this obscure physical entity, we shall deduce at first from relativistic electrodynamics expressed by means of points in space-time, the dynamics of an atomic or molecular system of any number of degrees of freedom. We shall then devise a general method of quantization in space-time, which we shall apply to the quantization of the point electron and to that of \textit{continuous} systems: It will be shown that this quantization is a logical consequence of our gravific theory [...]'\footnote{De Donder used the old term `gravific theory' instead of `gravitational theory'.} (\cite{DeDonder2}; p. 8). 

This comment is important for two reasons. First, it emphasized again that the problem of reconciling quantum physics and GR was considered early in the history of quantum physics. Secondly, De Donder developed his approach during the birth of QM and it is a ``spurious'' approach in the following sense. Before 1925 the quantization of a system was performed using Epstein-Sommerfeld-Wilson rules and a system like `the point electron', as De Donder referred to, would follow a classical trajectory. He agreed with this interpretation and in this sense, from our point of view, his approach belongs to the old quantum theory. But De Donder knew Schr\"odinger papers and he explicitly stated that he was looking for new quantization rules that should be compatible with the curved space-time of Einstein theory. These rules would have to reproduce, in his opinion, the general relativistic generalization of Schr\"odinger's equation\footnote{Once again the reference was to the KG equation.}. This means that with the phrase `general method of quantization in space-time' De Donder intended  a procedure to obtain a wave equation for the wave function $ \psi $, living on a curved background. As far as we know, De Donder never referred to $  \psi$ as a field. For this reason we could say that De Donder was looking for a ``General Relativistic Quantum Mechanics'' (GRQM). 

In WM a key ingredient of the quantization procedure was the imposition of boundary conditions for the wave function. As far as we know, De Donder never considered any boundary conditions explicitly. As we will see, his method was based on a unified variational principle, but De Donder's $ \psi $ was treated, from our point of view, classically. This means also that, from the modern field theoretic point of view, he did not consider any quantum feature of the fields. Lastly, it is worth noting that De Donder was not alone in believing that quantization rules could be derived in the context of some unknown classical theory. Einstein, for example, would look for a classical field theory (Einheitliche Feldtheorie) for the rest of his life \cite{Pais}. We do not know why De Donder was convinced of this idea, but because of the absence of a discussion on the wave function's boundary conditions, as we shall discuss after equation (\ref{quantDD}), the unified variational principle seemed not to require any modification of GR. For this reason, De Donder thought that the quantization rules should have been a consequence of GR principles, as he stated in the in the introduction cited above. This attitude is consistent with the claim that De Donder belongs to the group of authors who were convinced of GR supremacy. This conviction is confirmed by the last sentence of the general introduction to his MIT lectures: `Once more relativity unfolds the great physical drama of the universe clad in an immutable form bearing the stamp of eternal laws.' (\cite{DeDonder2}; p. 8). This means also that from a modern point of view, in his approach De Donder did not consider any quantum effect on the gravitational field. This fact was common to almost all the pre-1930 works: as far as we know Rosenfeld's approach was the only exception.        

We introduce some technical details in order to understand how De Donder tried to harmonize WM with GR. The tenth lecture is dedicated to the `Relativistic Quantization', and it started from the classical dynamics of a charged particle in GR, i.e. the `point-electron'. The dynamics is described by the Euler-Lagrange equations obtained using the following Lagrangian\footnote{The ``Lagrangian'' used by De Donder had the dimensions of a Lagrangian divided by a velocity and the same happens for the following ``Hamiltonian'' (\ref{hamil-DD}), but we will call them Lagrangian and Hamiltonian as well.} (\cite{DeDonder2}; p. 90):
\begin{equation}\label{az-DD}
L_{DD} \left( x ;\, u \right) = \frac{mc}{2}g_{\mu\nu} u^{\mu}u^{\nu}-\frac{e}{c}A_{\mu}u^{\mu} ,
\end{equation}
where $ \displaystyle  u^{\mu}= \,\frac{dx^{\mu}}{d\tau} $, $ \tau $ is the proper time, and the tangent vector satisfies the following constraint:
\begin{equation}\label{constraint}
g_{\mu\nu}u^{\mu}u^{\nu}=-1 .
\end{equation}
Using $ L_{DD} $, De Donder was able to define the conjugate momenta as  $\displaystyle p_\mu=\frac{\partial L_{DD}}{\partial u^\mu}=mcg_{\mu\nu}u^{\nu}-\frac{e}{c}A_\mu $, and the Hamiltonian $ H =p_\mu u^\mu-L_{DD} $ reads:
\begin{equation}\label{hamil-DD}
H = \frac{1}{2mc}\left( p_\mu +\frac{e}{c}A_\mu\right)\left( p^\mu +\frac{e}{c}A^\mu\right) \, .
\end{equation}
The constraint $ g_{\mu\nu}u^{\mu}u^{\nu}=-1 $ is equivalent to the relation $H =-\frac{1}{2}mc $, i.e. the reduced HJ equation for a point particle, which De Donder called `Jacobian equation'. Finally, by using equation (\ref{hamil-DD}), the constraint assumes the following form (\cite{DeDonder2}; p. 91, equation (10)):
\begin{equation}\label{eq-cl}
g^{\mu\nu}\left( \frac{\partial S}{\partial x^\mu} +\frac{e}{c}A_\mu\right) \left( \frac{\partial S}{\partial x^\nu} +\frac{e}{c}A_\nu\right) +m^2c^2 =0 \qquad ,\qquad  \frac{\partial S}{\partial x^\mu}=p_\mu,
\end{equation} 
where $ S $ is the Jacobi function of classical mechanics. Before going on, we point out that De Donder was aware of the following fact. Using $\displaystyle S_4 = -mc\int_O^M d\tau $ as action for the free point-particle, the Lagrangian approach could be performed introducing an arbitrary parameter $ \lambda $ and rewriting $ S_4 $ as follows:
\begin{equation}
S_4 = \int_O^M L d\hat{\lambda} = \int_O^M \sqrt{-\gamma_{\bar{\mu}\bar{\nu}}\frac{dx^{\bar{\mu}}}{d\hat{\lambda}}\frac{dx^{\bar{\nu}}}{d\hat{\lambda}}}\, d\hat{\lambda}\; .
\end{equation}
In this case, a Legendre transform would produce a null Hamiltonian, i.e. the constraint $ H=0 $.

At this point De Donder introduced a wave function associated to the electron, namely $ \psi\left( \tau, x\right)  $, a function of the spatial coordinates $ x $ and of the proper time $ \tau $. In the MIT lectures, the author made no explicit discussion neither on the mathematical feature of the wave function nor on its physical interpretation. He implicitly identified it with Schr\"odinger's wave function, when considering a single electron. In fact, De Donder imposed the following Ansatz for the wave function (\cite{DD1}; p. 91):
\begin{equation}\label{DD1}
\psi =e^{k\, S} \qquad \text{i.e.}\qquad  S=\frac{1}{k}\, \text{log}\left( \psi\right) \, ,
\end{equation}
where the Jacobi function $ S(\tau\, , x ) $ depends on the spatial coordinates and on the proper time. At the beginning $ k $ is an unknown constant, but in the end, in order to match his wave equation with Schr\"odinger equation, he would choose $ \displaystyle k = \frac{i}{\hbar} $. De Donder made no comment on the fact that with this choice both $ \psi $ and the log-function in equation (\ref{DD1}) turn into complex functions. As a consequence of the fact that he left $ k $ undetermined, he would not use the complex conjugate as we shall do in equation (\ref{functional}). De Donder will use the correct notation in his book on Variational Calculus \cite{DD-var}. If $ \displaystyle k = \frac{i}{\hbar} $, the Ansatz (\ref{DD1}) corresponds to the correct geometrical optics approximation. It is worth noting that this procedure is very similar to Klein's approach. In fact, this procedure was the common way to introduce a wave equation for a ``quantum'' particle in the mid 1920s. Unlike Klein, from De Donder's point of view it was not necessary to unify all forces with a five-dimensional Lagrangian. Indeed, De Donder was satisfied with a unified action principle. Unlike Klein, he looked from the beginning for an action principle in four dimensions, with the help of relativistic Hamiltonian dynamics. 

After having introduced the Jacobi function $ S(\tau\, , x ) $, in order to obtain the reduced HJ equation $H =-\frac{1}{2}mc $, the reducibility condition reads:
\begin{equation}\label{DD2}
\frac{\partial S}{\partial \tau}=\frac{1}{2}mc  .
\end{equation}
Integrating (\ref{DD2}), De Donder wrote the Jacobi function in the following form:
\begin{equation}\label{JacobiDD}
S = \frac{1}{2}mc\tau + S_0\left( x^0,\, x^1,\, x^2,\, x^3\right) \, ,
\end{equation}
that will play an important role for Rosenfeld, as we shall see in the next section.

Thanks to definition (\ref{DD1}) and using equation (\ref{DD2}), the author was able to write (\cite{DD1}; p.91):
\begin{eqnarray}
\frac{\partial S}{\partial \tau}&=&\frac{\hbar}{i}\frac{1}{\psi}\frac{\partial \psi}{\partial \tau}\, ,\label{DD3a}\\
\frac{\partial S}{\partial x^\mu}&=&\frac{\hbar}{i}\frac{\partial_\mu\psi}{\psi} \, ,\label{DD3b}\\
\psi &=& \frac{\hbar}{i}\frac{\frac{\partial \psi}{\partial \tau}}{\frac{\partial S}{\partial \tau}} = \frac{\hbar}{i}\frac{2}{mc}\frac{\partial \psi}{\partial \tau} \, .\label{DD3c}
\end{eqnarray}
The conjugated wave function $ \overline{\psi} $ satisfies the conjugated version of equations (\ref{DD3a}), (\ref{DD3b}) and (\ref{DD3c}).

Inserting (\ref{DD3b}) and (\ref{DD3c}) into (\ref{eq-cl}), the HJ equation (\ref{eq-cl}) can be rewritten in the following form:
\begin{equation}\label{functional}
 J\left(\psi \right) \equiv -g^{\mu\nu}\left( \frac{mc}{2}\partial_\mu\psi +\frac{e}{c}A_\mu\frac{\partial \psi}{\partial \tau}\right)\left( \frac{mc}{2}\partial_{\nu}\overline{\psi} -\frac{e}{c}A_{\nu}\frac{\partial \overline{\psi}}{\partial \tau}\right) - m^2c^2\frac{\partial \psi}{\partial \tau}\frac{\partial\overline{\psi}}{\partial \tau}=0\, .
\end{equation}
In De Donder's approach equation (\ref{functional}) defines a functional $  J\left(\psi \right) $, that is an invariant under all changes of variables, $ x^0, \dots ,x^3 $ (\cite{DeDonder2}; p. 92).
The $ J $ functional plays a fundamental role for the author. From his point of view, with the introduction of the wave function $ \psi $, the classical HJ equation (\ref{eq-cl}) becomes a constraint for the new functional $ J(\psi ) $, i.e. 
\begin{equation}\label{eq-Jac} 
 J\left(\psi \right) = 0 \, ,
\end{equation} 
and using this new functional De Donder was able to introduce what the author calls the relativistic quantization rule for curved space-time. After defining the following functional derivative:
\begin{equation}\label{der-funz}
\frac{\delta}{\delta\psi}J(\psi )=\frac{\partial J}{\partial \psi}-\partial_\mu\frac{\partial J}{\partial\partial_\mu\psi}+\dots \, ,
\end{equation}
the quantization rule reads: `the variational derivative of the left-hand member of the Jacobian equation (\ref{eq-Jac}), with respect to $ \psi $, shall vanish. Explicitly: 
\begin{equation}\label{quantDD} 
\frac{\delta}{\delta\psi}\left( \sqrt{-g} J\right) = 0\; \text{'}
\end{equation}
(\cite{DeDonder2}; p. 92).

Before going on, let us consider De Donder's variational principles in more detail. Lecture 5 of the MIT lectures is dedicated to `The fundamental Equations of the Gravific Field'. In order to obtain Einstein equations, De Donder considered the following variational principle (\cite{DD1}; p. 47):
\begin{equation}
\frac{\delta\left[\left(  aR+b + \mathcal{L}_m \right)\sqrt{-g}\right]  }{\delta g^{\mu\nu}} = 0\; ,
\end{equation}
where the functional derivative is defined as in equation (\ref{der-funz}) with $ \psi $ replaced by the metric, $ R $ is the four-dimensional curvature scalar, $ a $ and $ b $ are arbitrary constants (incidentally, the constant $ b $ plays the role of the Cosmological Constant $ \Lambda $, but De Donder did not comment on this fact), $ \mathcal{L}_m $ is an unspecified Lagrangian density for the matter part of the theory, and the functional $ \left(  aR+b \right)\sqrt{-g} $, i.e. the Lagrangian density, is named `\textit{the characteristic gravific function}' (\cite{DD1}; p. 47). It seems that in these years De Donder preferred to introduce a variational principle using Lagrangian densities instead of action functionals. De Donder himself stressed this fact as follows, advocating a precise justification of the choice he made: `The variational principle, as we have presented it, is evidently a generalization of Hamilton's principle, that is, equivalent to placing
\begin{equation}
\delta\int_{\Omega}\left(  aR+b + \mathcal{L}_m \right)\sqrt{-g} d^4x = 0\; ,
\end{equation}
$ \Omega $ being a region of space-time at the boundaries of which the variations must vanish. \textit{It is in order to avoid the use of four-dimensional space that we have preferred the above presentation}.' [emphasis added] (\cite{DD1}; p. 47). In his following works devoted on the developments of variational principles and their applications \cite{DD-1930}, the author will use both forms. Let us now consider again De Donder's approach to quantization procedure.

Why did De Donder call equation (\ref{quantDD}) `a quantization rule'? The functional derivative (\ref{der-funz}), introduced by De Donder, produces the usual equations of motion for a charged scalar field and he showed that it reduces to the Schr\"odinger's equation in the non relativistic limit and in the approximation of an electrostatic field. It is worth noting that De Donder's $ \psi $ would not have the correct dimensionality to be interpreted as the Schr\"odinger's wave equation, but De Donder made no comments on this fact. For this reason he considered equation (\ref{quantDD}) as a quantization rule. In this sense, for us, De Donder's approach belongs to the WM point of view: like Klein he believed that writing a wave equation was a sufficient condition to describe the quantum behaviour of a system.

Why did De Donder assert in his general introduction that this quantization rule would be `a logical consequence of our gravitational theory'? In order to answer this question, firstly we note that from a modern point of view, De Donder's approach is of course a classical approach, because it is equivalent to a classical variational principle for a field theory, though De Donder interpreted the ``field'' $ \psi $ as a wave function. The absence of the integral in (\ref{quantDD}) was compensated by an ad hoc choice of the functional derivative defined in (\ref{der-funz}). Secondly we remember that the first authors that tried to quantize scalar fields were Klein and Jordan in 1927 \cite{Klein4}. This means that the concept of quantum field was not already born and like other authors De Donder was convinced that writing a wave equation for a system was sufficient to quantize it. De Donder was convinced that GR could explain where the quantization rules come from, because he obtained Schr\"odinger's wave equation through the use of a variational principle, like Einstein's equations are obtained, only from different action. Lastly, it is worth noting that by applying variational methods without imposing commutation relations for the fields, the apparatus of GR seems not to require any modification. For these reasons, De Donder made the following remark, in order to emphasize his interpretation of the approach: `We have thus shown that \textit{the quantization of the point electron can be deduced from Einstein's gravitational theory} by means of an absolute extremal.' (\cite{DeDonder2}; p. 95).   
  
Before going on, we make the following remark on De Donder's functional. Unlike Klein, who considered a real scalar field in five dimensions, De Donder wrote a sort of Lagrangian density for a charged scalar field. More precisely, using relation (\ref{DD3c}) the $ J $ functional reads:
\begin{equation}\label{lagr-DD}
J(\psi) = \frac{m^{2}c^{2}}{4}\left[- g^{\mu\nu}\left( \partial_{\mu}\psi+\frac{i}{\hbar}\frac{e}{c}A_\mu\psi\right) \left( \partial_{\nu}\overline{\psi}-\frac{i}{\hbar}\frac{e}{c}A_\nu\overline{\psi}\right)-\frac{m^2c^2}{\hbar^2}\overline{\psi}\psi\right]
\, .
\end{equation}
The expression in the squared brackets resembles the Lagrangian density of a complex scalar field in the presence of an electromagnetic and a gravitational field, but neither $ \psi $ nor $ J $ would have the correct dimensionality to be interpreted as a scalar field and a density Lagrangian respectively. Unlike Klein's functional, De Donder's functional (\ref{lagr-DD}) would have the correct sign in order to be interpreted as a Lagrangian density \cite{Rocci1}.

\section{Rosenfeld's contributions}\label{Rosenfeld}
Rosenfeld merged De Donder's and de Broglie's ideas using Klein's approach. He explicitly cited all the authors we discussed in the preceding section. Like De Donder, he considered the relativistic Jacobi function approach. Like de Broglie, he explicitly inserted a mass term in the KG equation. Like Klein, he was aware of the fact that the fifth dimension's character should be space-like. But the principal purpose of Rosenfeld was to try to understand concretely how quantum effects should modify the classical view in the presence of a gravitational field, at least in the weak field approximation. 

All of Rosenfeld's papers on this topic, \cite{Ros1} \cite{Ros2} \cite{Ros3}, are authored by Rosenfeld alone: to what extent were de Broglie and De Donder active collaborators in these articles? The influence of de Broglie and De Donder is stated explicitly by the author himself. At the end of the introduction of his first paper, Rosenfeld wrote: `This work was completed under the direction of Mr. L. de Broglie and Mr. Th. De Donder, who have never ceased to assist me with their advice, and have been kind enough to communicate to me their works, even manuscripts; I am happy to be able to express my deep appreciation to them here.' (\cite{Ros1}; p. 305). From the observations that we make in the rest of this paper, we can infer that De Donder had an active part in Rosenfeld's paper. In particular, we shall see how Rosenfeld followed De Donder's approach to introduce the wave equation in the context of a curved space-time, which permitted him to find a natural explanation of De Donder's interpretation of the quantum wave amplitude. Furthermore, we shall infer what precisely de Donder found attractive in Rosenfeld's five-dimensional Universe. In his second and third communications, Rosenfeld supported with a physical explanation his first paper. Stimulated by De Donder's influence, Rosenfeld recognized that he was using Bohr's correspondence principle. Unlike Rosenfeld, De Donder thought that Rosenfeld's work was a proof of a new version of the correspondence principle, which could be derived from Einstein's theory, and stressed that this principle should have been a cornerstone or the `gravitational wave mechanics' (\cite{DD-Ros}; p. 506), i.e. a theory reconciling WM with Einstein's theory. 

Rosenfeld's first paper \cite{Ros1} is a long and technical work and it does not contain any physical interpretation of the choices he made. For this reason, in subsection \ref{calculation} we shall pay more attention to the technical details of the Rosenfeld's approach, explaining his results from the author's point of view. The second and the third papers are shorter than his first contribution. In these articles the author clarified his technical choices from the physical point of view. We will analyse Rosenfeld's comments in subsection\footnote{The fourth of Rosenfeld's communication is an attempt to unify the preceding works.} \ref{corr-princ}. At the end, in subsection \ref{discussion}, we shall emphasize how these first articles influenced Rosenfeld's future work and we shall interpret the author's results from a modern point of view.

\subsection{The quantum origin of a space-time metric}\label{calculation}
In the introduction to his first paper \cite{Ros1}, written during his stay in Paris at the ``\`Ecole normale sup\'erieure'', Rosenfeld formulated his main goals\footnote{We present an English translation of some parts of the original paper, written in French, and then we comment on it. We omit the references of the original work.}: 
\begin{quote}
`The first part of this work is dedicated to the systematic study of the five-dimensional universe considered by O. Klein, Th. De Donder and L. de Broglie. We will show how the model of the five-dimensional universe is satisfactory [...]. Generalizing Gordon's and Schr\"odinger's papers, we will show how the introduction of the $ \Psi $ function of de Broglie-Schr\"odinger permits us to combine in a unique variational principle, into the five-dimensional universe, the gravitational force, the electromagnetic force and the quantum phenomena (the $ \Psi $ equation). [...] \textit{Finally, a formula will be established to calculate the gravitational and electromagnetic potentials, for a field slightly different from the Minkowskian field, as a function of} $ \Psi $. The calculation will be developed for the case of a stationary charge and for the case of a charge moving with constant speed. Comparing the values obtained with the classical potentials, we find that the amplitude of the $ \Psi $ function representing the charge must have a constant value inside a finite volume and it must be zero outside of that volume: these results can be well understood with the beautiful interpretation of the  $ \Psi $ function recently proposed by Mr. De Donder; quite to the contrary it appears to be irreconcilable with the opinion of Mr. de Broglie, who believed that the charge would be a point singularity of the  $ \Psi $ function. [emphasis added]' (\cite{Ros1}; p. 304-5). 
\end{quote}
We shall investigate only the first case proposed by Rosenfeld, i.e. the case of a stationary massive charge, represented by a wave function, in order to investigate the gravitational field produced by a quantum particle. Rosenfeld would consider a weak-field approximation, what he called `a field slightly different from a Minkowskian field'\footnote{Minkowskian field is the English translation of the French expression ``champ de Minkowski'' which was well understood and commonly used in that period as the vacuum space. See e.g. \cite{Solomon} or Lichnerowitz in \cite{Pauli-letter}.}. Rosenfeld would find that the quantum particle should be represented by a localized wave function, which is non zero inside a finite volume, instead of a point-like object, in contrast with de Broglie's point of view. This fact would enforce De Donder's interpretation of the wave function's amplitude as representing a sort of internal quantum force of matter. We will not discuss this interpretation, which was based on the application of Rosenfeld-De Donder's approach to multi-particle systems, because for this case Rosenfeld did not investigate the gravitational field.

Why did Rosenfeld consider a five-dimensional framework? The answer seems now almost trivial: the author studied Klein's work with de Broglie and was fascinated by its capability to describe in a unified framework GR and Maxwell's theory. 

What was Rosenfeld's starting point? The answer is connected with his knowledge of De Donder's and de Broglie's works. Indeed, following De Donder, Rosenfeld started from the classical description of a single charged particle, and following Klein and de Broglie, he considered a five-dimensional space-time, with the usual coordinates $( x^0\, ,\, x^1\, ,\, x^2\, ,\, x^3\, ,\, x^5\, )  $. The classical particle was described by a five-dimensional Jacobi function $ \bar{S} $, namely
\begin{equation}\label{S5-Ros}
\bar{S}\left( x \right) =-\frac{e}{c\beta}x^5 + S_0\left( x^0\, ,x^1\, , x^2\, , x^3 \right)\, ,
\end{equation} 
in analogy with De Donder's four-dimensional Jacobi function (\ref{JacobiDD}), that we rewrite here for convenience, namely:
\begin{equation}\label{JacobiDD1}
S = \frac{1}{2}mc\tau + S_0\left( x^0,\, x^1,\, x^2,\, x^3\right) \, .
\end{equation}
Rosenfeld explicitly defined the fifth coordinate putting:
\begin{equation}\label{5d-Ros}
\text{`}x^5 = -\frac{mc^2\beta}{2e}\tau\, \text{' (\cite{Ros1}; equation (5) p. 306),}
\end{equation} 
specifying that `$ \beta $ is a \textit{universal constant}.' (\cite{Ros1}; p. 306). From our point of view, the introduction of the fifth coordinate simply follows from the comparison between De Donder's Jacobi function,  equation (\ref{JacobiDD1}), and de Broglie's five-dimensional Hamiltonian action for the charged particle, equation (\ref{action-dB1}). Indeed, to obtain equation (\ref{action-dB1}), it is sufficient in (\ref{JacobiDD1}) to set $\displaystyle S_0 =  - \int_O^M\frac{e}{c}A_{\mu}dx^\mu -mc\int_O^M d\tau $. About the size of the fifth dimension, Rosenfeld shared de Broglie's view. He observed that from equation (\ref{S5-Ros}) it follows the invariance of $ x^5 $ with respect to the general transformation of coordinates $ f(x^0,\, x^1,\, x^2, \, x^3) $ and concluded: `Its invariance with respect to the transformations that we are able to perform explains why this fifth dimension escapes direct observations.' (\cite{Ros1}; p. 307). Like de Broglie, Rosenfeld did never discuss explicitly the size of the fifth dimension, though he would have been able to extract it\footnote{See the discussion after equation (\ref{tir}).}.

The dynamics of classical charged particles is described by the HJ equation and Rosenfeld introduced his five-dimensional analogously. Following the author we note first that the new Jacobi function $ \bar{S} $ satisfies\footnote{Note that the combination $ \frac{e}{c\beta}x^5 $ has the dimension of an action.}
\begin{equation}\label{eq7}
\partial_5\bar{S}=-\frac{e}{c\beta}\, .
\end{equation}
Secondly, Rosenfeld used Klein's five-dimensional metric $ \gamma_{\bar\mu\bar\nu} $ defined in the previous section, see equations (\ref{nota-d-sigma1b}) and (\ref{nota-d-sigma2b}), with the same convention, i.e. imposing the following choice for $ \alpha $ and $ \beta $: $ \alpha\beta^2 = 2\kappa $. Lastly, with the help of the components of the inverse metric $ \gamma^{\bar{\mu}\bar{\nu}} $, namely
\begin{equation}\label{inverse}
	\gamma^{\mu\nu} = g^{\mu\nu}\; ,\qquad \gamma^{55} = \frac{1}{\alpha}+\beta^{2}A_\mu A^\mu\; ,\qquad \gamma^{5\mu}= -\beta A^{\mu}\; ,
\end{equation}
the author is able to show how De Donder's four-dimensional HJ equation (\ref{eq-cl}), namely
\begin{equation}\label{HJ-Ros1}
g^{\mu\nu}\left(\partial_\mu S_0 + \frac{e}{c}A_\mu\right) \left( \partial_\nu S_0 +\frac{e}{c}A_\nu\right) +m^2c^2 =0\; ,
\end{equation} can be rewritten in the following compact form (\cite{Ros1}; p. 307):
\begin{equation}\label{eq-cl2}
\gamma^{\bar{\mu}\bar{\nu}}\partial_{\bar{\mu}}\bar{S}\partial_{\bar{\nu}}\bar{S} =
-\left( m^2c^2 - \frac{e^2c^2}{16\pi G} \right)\, .
\end{equation}
It is worth noting that equation (\ref{eq7}) is the same relation that induced Klein to introduce a fifth coordinate : it suggests indeed that the electric charge could play the role of an extra momentum component, as recollected by Klein (see the beginning of section \ref{section-Klein}), and permits to translate in the five-dimensional language the relativistic HJ equation for a particle moving in a combined electromagnetic and gravitational field. 

Choosing $ \alpha\beta^2 = 2\kappa $, Rosenfeld implicitly imposed $ \alpha > 0 $. As noted in the previous section, this means that, like Klein, Rosenfeld correctly introduced a space-like fifth dimension. Hence, the quantity $ \mathcal{I}^{2} $, see equation (\ref{Iquadro}), assumes the following form: 
\begin{equation}\label{Iquadro-Ros}
 \mathcal{I}^2_{Ros} = m^2c^2 - \frac{e^2c^2}{16\pi G}\; ,
\end{equation}
and it differs from de Broglie's $ \mathcal{I}_{dB} $, see equation (\ref{Iquadro-dB}), because of the presence of the minus sign. For an electron, the quantity $ \mathcal{I}^2_{Ros} $ is negative: indeed Rosenfeld did not use the symbol $ \mathcal{I}^{2}_{Ros} $, but he explicitly wrote its square root, cfr. eq. (\ref{RosA}) below. Hence, we introduced it in order to compare Rosenfeld's and de Broglie's work. As we shall see in a moment, Rosenfeld did not discuss the square root of the expression $ \mathcal{I}_{Ros} $, but he underlined that it has a geometrical meaning as follows. Parametrizing the five-dimensional path with $ \hat{\tau} $ and the particle's four-dimensional world line with the proper time $ \tau $, Rosenfeld wrote: `It is easy to calculate the five-dimensional trajectory's slope on the space-time. Indeed, if $ \bar{S} $ is a complete integral of equation (\ref{eq-cl2}), along the trajectory, from (\ref{eq-cl2}) it follows that
\begin{equation}\label{RosA}
\gamma^{\bar{\mu}\bar{\nu}}\partial_{\bar{\nu}}\bar{S}=\sqrt{m^2c^2 - \frac{e^2c^2}{16\pi G}}\cdot\frac{dx^{\bar{\mu}}}{d\hat{\tau}}\, ,
\end{equation}
and from (\ref{eq7}), (\ref{HJ-Ros1}) and (\ref{inverse}) it follows that
\begin{equation}\label{RosB}
\gamma^{\mu\bar{\nu}}\partial_{\bar{\nu}}\bar{S}=mc\frac{dx^{\mu}}{d\tau}\, .
\end{equation}
This means that the slope reads:
\begin{equation}\label{slope}
\frac{d\hat{\tau}}{d\tau}=\sqrt{1-\frac{1}{2\kappa\mu^2}}
\end{equation}
and therefore it is determined only by the ratio $ \mu $; this geometric interpretation of the ratio $ \mu $ was \textit{on the ground of de Broglie's reasoning}.'\footnote{See \cite{Landau-teo-campi} for an explanation of the four-dimensional case. Inserting equation (\ref{RosA}) into (\ref{eq-cl2}), it can be verified that (\ref{RosA}) is a complete integral of (\ref{eq-cl2}).} [emphasis added] (\cite{Ros1}; p. 308). The ratio $ \mu $ is defined by $ \mu=-\frac{mc^2}{e} $ and it encodes the characteristics of the particle, because it involves the particle's mass and charge. The emphasis added at the end of the citation underscores de Broglie's influence on Rosenfeld's approach. Firstly, Rosenfeld's equation (\ref{slope}) is equivalent to de Broglie's equation (\ref{relazioni}). Secondly, in the previous section we said that from de Broglie's point of view $ P_{\bar{\nu}}=\partial_{\bar{\nu}}\bar{S}  $ should be interpreted as the five-dimensional generalization of $ p_\mu = mcg_{\mu\nu}\frac{dx^{\nu}}{d\tau} $. Rosenfeld referred to the fact that equations (\ref{RosA}) and (\ref{RosB}) made explicit this connection\footnote{In appendix \ref{app3a} we clarify the connection among equations (\ref{RosA}), (\ref{RosB}) and (\ref{slope}).}, because they implied that $ \gamma^{\mu\bar{\nu}}P_{\bar{\nu}}=g^{\mu\nu}p_\nu $. Furthermore, Rosenfeld agreed explicitly with de Broglie's idea that the particle's five-dimensional geodesics would be inclined with respect to the hyperplane that locally describes the four-dimensional hypersurface $ x^5 = const. $. See de Broglie's comments after equation (\ref{d-sigma3}).

After having introduced the five-dimensional Universe and its unified description of the gravitational and electromagnetic interaction, the author introduced what he called the `de Broglie-Schr\"odinger wave function' (\cite{Ros1}; p. 311). Following de Broglie and De Donder, equations (\ref{deB-optic}) and (\ref{DD1}), Rosenfeld's general Ansatz for the five-dimensional wave function reads:
\begin{equation}\label{fun-onda} 
\Psi \left( x \right)  =\mathcal{A}\left( x^0\, ,x^1\, , x^2\, , x^3 \right)  e^{k\,\bar{S}} \; ,
\end{equation}
where $ \bar{S} $ is the Jacobi function (\ref{S5-Ros}), $ k $ is a constant and the amplitude $\mathcal{A}$ is in general a complex function of the form $\mathcal{A}= A +iB $. Like De Donder, Rosenfeld made the choice $ k = \frac{i}{\hbar} $ and then he considered the case of real constant amplitude, in order to compare his five-dimensional functional with De Donder's $ J $ functional. But Rosenfeld assigned the value of $ k $ ab initio, therefore, as we pointed out in the discussion after equation (\ref{DD1}), both De Donder and Rosenfeld considered wave functions as complex objects. The periodicity condition is still contained in Rosenfeld's Ansatz (\ref{fun-onda}), because the wave function is periodic in the fifth coordinate, see equation (\ref{S5-Ros}). In the case of real constant amplitude $ A $, from equation (\ref{fun-onda}) it follows:
\begin{equation}
\frac{\partial \bar S}{\partial x^{\bar\mu}}=\frac{\hbar}{i}\frac{\partial_{\bar\mu}\Psi}{\Psi}\label{def-onda2}\, .
\end{equation}
Inserting (\ref{def-onda2}) into the HJ equation (\ref{eq-cl2}), Rosenfeld obtained the five-dimensional generalization of De Donder's functional equation (\ref{eq-Jac}), i.e. $ \displaystyle \mathcal{L} = 0 $, where the new functional is
\begin{equation}\label{lagr-Ros} 
\mathcal{L}\left(\,\Psi \, ,\overline{\Psi}\,\right) = - \gamma^{\bar{\mu}\bar{\nu}}\partial_{\bar{\mu}}\overline{\Psi}\partial_{\bar{\nu}}\Psi-
\frac{\mathcal{I}^{2}_{Ros} }{\hbar^2}\overline{\Psi}\Psi\; ,
\end{equation}
the symbol $ \overline{\Psi} $ is the complex conjugate of the five-dimensional wave function and we used for this quantity the symbol $ \mathcal{I}_{Ros} $, equation (\ref{Iquadro-Ros}), for brevity. This means that from Rosenfeld's point of view the constant amplitude case corresponded to the classical limit. Indeed, the author underlined: `In the general  case, i.e. when $ \mathcal{A} $ is an arbitrary function, $ \mathcal{L} $ is no longer null along a trajectory.' (\cite{Ros1}; p. 312). As a consequence $ \mathcal{L} $ is able to play a central role for the quantum dynamics.   

Following De Donder, the quantum picture would be described by a variational principle involving (\ref{lagr-Ros}): Rosenfeld applied De Donder's functional derivative (\ref{der-funz}) on $ \mathcal{L}\sqrt{-g} $ and obtained, by varying with respect to $ \overline{\Psi} $ and $ \Psi $ independently, the following wave equations:
\begin{equation}\label{eomRos}
\gamma^{\bar{\mu}\bar{\nu}}\nabla_{\bar{\mu}}\partial_{\bar{\nu}}\Psi =\frac{\mathcal{I}^{2}_{Ros}}{\hbar^2} \Psi \qquad\quad\text{and}\quad\qquad
\gamma^{\bar{\mu}\bar{\nu}}\nabla_{\bar{\mu}}\partial_{\bar{\nu}}\overline{\Psi} =\frac{\mathcal{I}^{2}_{Ros} }{\hbar^2} \overline{\Psi}\, ,
\end{equation}
and that should be, as Rosenfeld wrote, `a generalization of the de Broglie-Schr\"odinger's equation' (\cite{Ros1}; p. 312), i.e. equation (\ref{KG4}). Having introduced a complex wave function ab initio, Rosenfeld wrote explicitly a wave equation both for $ \Psi $ and for $ \overline{\Psi} $. The author's functional $ \mathcal{L} $ is formally equivalent to the Lagrangian density of a complex scalar field, but as for all of the authors of this period, $ \Psi $ is treated as a wave function. This approach has been conceived in a period that lies between the birth of QM and the birth of QFT, when scholars were looking for a ``relativistic quantum mechanics''. For this reason we could say that, like De Donder, Rosenfeld was looking for GRQM. The wave equation obtained by varying $ \overline{\Psi} $ in (\ref{lagr-Ros}) is formally equivalent to the five-dimensional wave equation suggested by de Broglie (\ref{KG4}). Rosenfeld used De Donder's variational derivative, but he was aware of the fact that this procedure is equivalent to the variational principle used in a modern field theory, obtained varying the integral of the Lagrangian density and imposing that the variations of the fields should be zero at the boundary of the domain of integration. Indeed, Rosenfeld claimed that $ \mathcal{L} $ should be the generalization of the Lagrangian considered by Gordon in \cite{Gordon}, where Gordon himself suggested to consider the wave function and his complex conjugated as independent variables with vanishing variations at the boundary. Unlike Klein's functional, Rosenfeld's $ \mathcal{L} $ functional had the correct sign to be interpreted as a Lagrangian density \cite{Rocci1}. This follows from the fact that Rosenfeld was influenced by De Donder's approach presented above. Unlike De Donder, Rosenfeld considered a general form for the wave functions, admitting that its amplitude $ A $ could be a non-constant function of the four-dimensional coordinates. Rosenfeld noted that in the constant-amplitude case he obtained De Donder's results, which are connected with the classical HJ equations (\ref{eq-cl2}) as suggested by De Donder himself. 

How did Rosenfeld reconcile GR with QM? Like De Donder, after having used the wave-particle duality via the Hamiltonian dynamics, Rosenfeld supposed that, in the case of non-constant amplitude, $ \mathcal{L}$ should be the correct generalization of Schr\"odinger's Lagrangian \cite{Schroe2} in the sense of GRQM. Finally, Rosenfeld introduced a variational principle, based on the following five-dimensional action\footnote{In equation (\ref{az-glob2}) the determinant of the four-dimensional metric $ g $ appears, instead of $ \gamma $. In Rosenfeld's approach, the two determinants are related by the relation  $ \gamma = \alpha g $ as explained in appendix \ref{app3b}. This means that the presence of $ g $ does not affect the equations obtained by varying (\ref{az-glob2}).} 
\begin{equation}\label{az-glob2}
\mathcal{S}_{tot}\left(\, \gamma\, , \Psi\, ,\overline{\Psi}\,\right) =\int d^5x  \sqrt{-g}\left[ -\tilde{R} +2\kappa\mathcal{L} \right]  \, ,
\end{equation}
where $ 2\kappa = \frac{16\pi G}{c^{4}} $. Rosenfeld did not specify the domain of integration, we suppose that the integral should be performed over an arbitrary portion $ \Omega $ of the five-dimensional space-time. By varying the action with respect to the metric like in equation (\ref{variation}), he obtained the five-dimensional Einstein equations coupled with the complex field $ \Psi $, which are formally equivalent to a system with the four-dimensional Maxwell equations coupled to the scalar field and the four-dimensional Einstein equations coupled to the electromagnetic and the scalar fields. By varying the action with respect to $ \overline{\Psi} $ and $ \Psi $, using De Donder's functional derivative, Rosenfeld obtained the KG equations (\ref{eomRos}) for $ \Psi $ and $ \overline{\Psi} $, respectively, as before, because the curvature's scalar depends neither on the wave function nor its complex conjugate. This is the unified framework that should reconcile, from Rosenfeld's point of view, GR with WM.

Did the five-dimensional formalism offer any additional insights beyond these that De Donder could have deduced in his four-dimensional context? As Rosenfeld stressed, the main advantage offered by the five-dimensional Universe was the opportunity to write a unified variational principle (\cite{Ros1}; p. 304). It is worth noting that the neutron would be discovered five years later \cite{Chadwick}. This means that all known elementary particles were charged particles and the unified picture offered by the five-dimensional Universe seemed to be a way to describe the known physical phenomena. As we shall see in subsection \ref{corr-princ}, Rosenfeld's approach permitted also to incorporate and, in a certain sense, to justify some of De Donder's ideas. 

It is not clear whether Rosenfeld considered his approach as a result or as a point of departure. But it is evident that he tried, for the first time, to investigate the geometry created by the wave function $ \Psi $. In fact, the equations obtained by varying action (\ref{az-glob2}) with respect to the metric are:       
\begin{equation}\label{eq-glob-5D}
\tilde{R}_{\bar{\mu}\nu}-\frac{1}{2}\gamma_{\bar{\mu}\nu}\tilde{R} = \kappa T_{\bar{\mu}\nu} \; ,
\end{equation}
where Einstein's and Maxwell's equations are coupled to the complex scalar field via the stress-energy tensor $ T_{\bar{\mu}\nu} $, defined by Rosenfeld as
\begin{equation}\label{def-T}
T_{\bar\mu\bar\nu} =  -\frac{2}{\sqrt{-g}}\frac{\delta \left(\sqrt{-g} \mathcal{L} \right) }{\delta \gamma^{\bar\mu\bar\nu}}\; ,
\end{equation}
which has the usual form:
\begin{equation}\label{tens-psi1}
T_{\bar{\mu}\nu}=\partial_{\bar{\mu}}\overline{\Psi}\partial_{\nu}\Psi +\partial_{\nu}\overline{\Psi}\partial_{\bar{\mu}}\Psi + \gamma_{\bar{\mu}\nu} \mathcal{L}  \; .
\end{equation}
Rosenfeld made no comments on the fact that in general the r.h.s. of equation (\ref{eq-glob-5D}) is a complex quantity. It is worth noting that the author investigated a particular case, i.e. when the wave function's amplitude is real. Hence, the energy momentum tensor is a real quantity. Introducing the wave function on the right side of equations  (\ref{eq-glob-5D}), Rosenfeld considered implicitly the wave function as representing the material part creating gravity. In this first paper, a long and technical paper, Rosenfeld did not justify this choice, which seems to be in contrast with the probabilistic interpretation of the wave function, from a modern point of view. As we shall see in the next section, the author would clarify his choice in the following work, where he referred explicitly to Bohr's correspondence principle. 

Like Klein, Rosenfeld did not consider the $ 55 $ component of the equations of motion: the Belgian physicist explicitly stated that this equation can be neglected, because the constancy of $ \gamma_{55} $ implies $ \delta\gamma_{55}=0 $ (\cite{Ros1}; p. 314)\footnote{As we said in the previous section, this is not correct.}. 

Before going on, we compare briefly Rosenfeld's approach with that of his mentors. Though Rosenfeld started out generalizing De Donder's approach, the unitary variational principle is presented starting with the action functional (\ref{az-glob2}) instead of De Donder's invariants, i.e. density Lagrangians. It is worth noting that in the same year Klein published independently a similar action, using a real scalar field. Klein coupled matter and geometry exactly like Rosenfeld did (\cite{Klein5}; p. 207). Unlike Rosenfeld, in \cite{Klein5}, Klein will express explicitly some perplexities about this kind of approach, observing that a unified action principle, e.g. that based on (\ref{az-glob2}), was only a starting step towards a unified theory that reconciles WM with GR (\cite{Klein5}; p. 190, footnote $ (^*) $ at the end of the introduction). In contrast, Rosenfeld, and De Donder with him, seemed to be convinced that the five-dimensional unified action principle would have some interesting features. Thanks to this conviction, the Belgian physicist investigated the quantum character of the metric produced by a quantum object, represented by the wave function $ \Psi $. 

In order to face this problem, Rosenfeld considered the weak-field approximation for the gravitational field, introduced by Einstein in 1916 to study the problem of gravitational waves, because it permitted to integrate the Einstein equations. In this approximation the metric can be written in the following form (\cite{Ros1}; p. 319):
\begin{equation}\label{metr-lin-a} 
\gamma_{\bar{\mu}\nu}=\eta_{\bar{\mu}\nu} + h_{\bar{\mu}\nu} \, , 
\end{equation}
where $ \eta_{\bar{\mu}\nu} $ is the five-dimensional Minkowski metric and $ h_{\bar{\mu}\nu} $ represents the perturbation of the flat metric, which satisfies the condition $ \vert h_{\bar{\mu}\nu} \vert << 1 $. Rosenfeld contracted (\ref{eq-glob-5D}) with $ \gamma^{\nu\bar{\mu}} $ to obtain an expression for the five-dimensional curvature scalar $ \tilde{R} $, namely\footnote{See appendix \ref{app3d} for a detailed explanation.}
\begin{equation}\label{formula-R-tilde}
\tilde{R} = -\kappa\left[ \gamma^{\nu\bar{\mu}}T_{\bar{\mu}\nu} + \frac{F_{\sigma\lambda} F^{\sigma\lambda}}{2}-\gamma^{\mu\rho}A_\rho\nabla_\lambda\left( \gamma_{\mu\sigma}F^{\sigma\lambda}\right) \right] \; .
\end{equation}
After having inserted (\ref{formula-R-tilde}) into equation (\ref{eq-glob-5D}), Rosenfeld used the Ansatz (\ref{metr-lin-a}) for the metric and he considered linear terms only obtaining:
\begin{eqnarray}\label{Einst-lin-5D}
\square h_{\bar{\mu}\nu} = -\kappa\left[ T_{\bar{\mu}\nu}-\frac{1}{2}\eta_{\bar{\mu}\nu} \eta^{\lambda\bar{\sigma}}T_{\bar{\sigma}\lambda} \right]\nonumber \\
= -\kappa \bar{T}_{\bar{\mu}\nu} \, ,
\end{eqnarray}
where the $ \square $ operator acts only on the usual four dimensions, because the metric does not depend on the fifth coordinate. In this approximation we are considering the gravitational field strength far away from the source, i.e. the particle's wave function, and the second and third term in the r.h.s. of equation (\ref{formula-R-tilde}) can be ignored in the case of a stationary charge\footnote{Rosenfeld did not write explicitly equation (\ref{Einst-lin-5D}), he referred to a `well known procedure' (\cite{Ros1}; p. 319) and wrote directly equation (\ref{sol-eq}).}. The stress-energy tensor appearing in (\ref{Einst-lin-5D}) has the same form of equation (\ref{tens-psi1}), but the curved metric $ \gamma_{\bar{\mu}\nu} $ has been substituted by the flat metric (\cite{Ros1}; p. 319). In particular, in this approximation the indices are raised and lowered by $ \eta_{\bar{\mu}\bar{\nu}} $. 
Rosenfeld was now able to integrate (\ref{Einst-lin-5D}), and obtained, using Rosenfeld's original notation\footnote{Rosenfeld did not specify that the integration is carried over a three-dimensional hypersurface $ \Sigma $ at the retarded time. In appendix \ref{app3e} we express equation (\ref{sol-eq}) in a modern notation. In the rest of our paper we will continue to use Rosenfeld's original notation.} (\cite{Ros1}; p. 319, equation (71)):
\begin{equation}\label{sol-eq}
h_{\bar{\mu}\nu} = -\frac{\kappa}{2\pi}\int\left\lbrace \bar{T}_{\bar{\mu}\nu}\right\rbrace_{t-\frac{r}{c}}\frac{dxdydz}{r}\; , 
\end{equation}
where, according to Rosenfeld, $ r $ represents the radial distance and the symbol $ \left\lbrace u\right\rbrace_{t-\frac{r}{c}}  $ means that the function $ u $ has been calculated using the variable $ t-\frac{r}{c} $: for this reason the (\ref{sol-eq}) components are often called retarded potentials. 
In order to consider the case of a stationary mass, the author chooses the following form\footnote{Remember that in our notation the combination $ \frac{e}{c\beta}x^{5} $ has the dimensions of an action.} for the Jacobi function $ \bar{S} $ (\cite{Ros1}; p. 320):
\begin{equation}\label{psi}
 \bar{S}= -\frac{e}{c\beta}x^{5} + mcx^{0}\, ,
\end{equation}
that appears in (\ref{fun-onda}), where now the amplitude $ A $ is a real function of the four-dimensional coordinates. Using this Ansatz, Rosenfeld was able to calculate explicitly the retarded potentials. Introducing the following functions\footnote{The integration domain is the same as in equation (\ref{sol-eq}).} of $  \mathbf{x} $ and $ t $:
\begin{eqnarray}
\mathcal{F} &=&\frac{2mc^2}{\hbar^2} \int \left\lbrace A^{2}\right\rbrace_{t-\frac{r}{c}} \frac{dxdydz}{r}\label{F}\; ,\\
\mathcal{W}_{\mu\nu} &=& \int \left\lbrace \partial_\mu A\partial_\nu A\right\rbrace_{t-\frac{r}{c}} \frac{dxdydz}{r}\; ,\\
\mathcal{G} &=& \int \left\lbrace \partial_\mu A\partial^{\mu} A\right\rbrace_{t-\frac{r}{c}} \frac{dxdydz}{r}\; ,
\end{eqnarray} 
the perturbations of the flat metric are therefore\footnote{In equation (\ref{metr-quant-d}) we used explicitly  that $ \alpha $ and $ \beta $ satisfy the constrain $ \alpha\beta^2 = 2\kappa $, like in Klein's approach.}:
\begin{eqnarray}
h_{5i} &=& 0\; , \qquad\qquad i = 1,2,3\; ,\label{metr-quant-a} \\
h_{50} &=& -\alpha\beta\left( \frac{e}{4\pi}\frac{\mathcal{F}}{c^2}\right) \; ,\label{metr-quant-b}  \\
h_{\mu\nu} &=& \frac{8G}{c^4}\mathcal{W}_{\mu\nu}\qquad\qquad \mu\neq \nu\; ,\label{metr-quant-c}\\
h_{\mu\mu} &=& \frac{2mG}{c^{4}}\mathcal{F} +\frac{8G}{c^4}\mathcal{G}\; .\label{metr-quant-d}
\end{eqnarray} 

It is worth noting that in (\ref{metr-quant-b}) and in (\ref{metr-quant-d}) the Planck constant appears via tha definition of $ \mathcal{F} $ (\ref{F}). In this sense, Rosenfeld's result represents a quantum correction of the flat metric. This is not surprising, because these corrections are generated by the wave function $ \Psi $. In this sense, the result is the first attempt to describe a quantum metric using WM and GR. As far as we know, this is the first time that a quantum metric appears in the history of QG. 

Rosenfeld did not emphasize this feature of the metric he found. As we have said, in his first paper Rosenfeld did not make explicit comments on the physical meaning of the calculations performed. As we shall see, in his following papers he would advocate Bohr's correspondence principle in explaining his use of the wave function as the source of gravitational field. From this perspective, it is easier to understand why Rosenfeld was more interested in analysing the metric in the case of a constant amplitude. Indeed, he considered a sort of semi-classical limit, confronting his ``quantum metric'' with its classical analogue. In this limit, equations (\ref{metr-quant-a}), (\ref{metr-quant-b}), (\ref{metr-quant-c}) and (\ref{metr-quant-d}) should match the metric produced by a classical source of mass $ m $ and charge $ e $, sitting at the origin \textbf{O} of the coordinates, at least in the weak-field limit, known today as the RN solution. The classical metric is presented in appendix \ref{app3f}, equation (\ref{RN}). At asymptotically large distances from the source it can be written as $ \gamma^{RN}_{\bar{\mu}\bar{\nu}} = \eta_{\bar{\mu}\bar{\nu}} + h^{RN}_{\bar{\mu}\bar{\nu}} $, where the components of the perturbations of the flat metric are:
\begin{eqnarray} 
h_{5i} &=& 0\; , \qquad i =1,\, 2,\, 3\; , \label{metr-cl-a}\\
h_{50} &=& \alpha\beta A_0 \quad\text{where}\quad A_0=\eta_{00}A^0=V=-\frac{e}{4\pi r_{0}}\; ,\label{metr-cl-b}\\
h_{\mu\nu} &=& 0\qquad \mu\neq \nu \; ,\label{metr-cl-c}\\
h_{\mu\mu} &=& \frac{2mG}{c^{2}r_{0}}\; ,\label{metr-cl-d}
\end{eqnarray}
where, according to Rosenfeld, $ r_0 $ represents `the distance between the origin \textbf{O} and an arbitrary point [of the five-dimensional space-time]' (\cite{Ros1}; p. 321). Equations (\ref{metr-cl-c}) and (\ref{metr-cl-d}) represent the components of the RN metric in the weak field approximation expressed using isotropic Cartesian coordinates\footnote{See appendix \ref{app3f} for a detailed discussion.}, while (\ref{metr-cl-a}) and (\ref{metr-cl-b}) coincide with $ \gamma_{5\mu} $ components (\ref{def-beta}) in  the case of a stationary charge. As we shall see in a moment, in considering the matching between classical metric and ``quantum metric'' in the semiclassical limit,  Rosenfeld did not consider a point-like charge, hence $ r_0=r_0(\vec{x}) $ should be a sort of ``mean distance'' from the charged body, sitting at the origin of the coordinates.

In order to match (\ref{metr-quant-a})-(\ref{metr-quant-d}) with (\ref{metr-cl-a})-(\ref{metr-cl-d}), $ \mathcal{W}_{\mu\nu} $ and $ \mathcal{G} $ must be zero and, as a consequence, the two following conditions must hold:
\begin{eqnarray} 
\partial_\mu A & = &  0 \quad , \label{cond-Ros-a}\\
\mathcal{F} & = & \frac{c^2}{r_0}\, . \label{cond-Ros-b}
\end{eqnarray}
Equation (\ref{cond-Ros-a}) follows directly from the condition $ \mathcal{W}_{\mu\nu}=0 $, while equation (\ref{cond-Ros-b}) can be obtained comparing (\ref{metr-cl-b}) with (\ref{metr-quant-b}). Rosenfeld discussed both these relations: `The first condition tells us that a fixed charge can be represented by a wave with \textit{stationary} phase and \textit{constant} amplitude.' (\cite{Ros1}; p. 322). As stated above, though Rosenfeld did not emphasize this fact, the constancy of the amplitude, i.e. condition (\ref{cond-Ros-b}), emerged as a condition to ensure that the quantum description could contain, at least as a limiting case, the classical description, which in this context corresponds to the classical five-dimensional RN metric (\ref{metr-cl-a})-(\ref{metr-cl-d}).  Besides this, the wave function of a fixed charge should have a fixed energy $ \mathcal{E} = mc^{2} $, and because of Heisenberg's uncertainty principle it should spread over the whole space. In a semi-classical approximation the wave packet is highly localized. Rosenfeld used a ``localized wave function'' instead, in the sense that Rosenfeld's wave function is non-zero only inside an arbitrary volume $ V $. Indeed Rosenfeld continued: `The second condition is satisfied [...] if we imagine that the amplitude is non-zero inside a finite volume centred around \textbf{O}.' (\cite{Ros1}; p. 322). Finally, using the \textit{mean value theorem}, the author defined formally the ``mean distance''\footnote{See appendix \ref{app3f} for a definition of the mean distance using modern notation.} $ r_{0} $ (\cite{Ros1}; p. 322):
\begin{equation}\label{def-r0} 
\frac{V}{r_0} =\int \frac{dxdydz}{r}\; .
\end{equation}
As usual, Rosenfeld did not specify the domain of integration. We suppose that it is the region where the wave function is non-zero, i.e. the volume $ V $.
By using definition (\ref{def-r0}) and the definition of $ \mathcal{F} $, equation (\ref{F}), in the constant amplitude approximation the condition (\ref{cond-Ros-b}) reads:
\begin{eqnarray}
\mathcal{F}=\frac{2mc^2}{\hbar^2} \int \left\lbrace A^{2}\right\rbrace_{t-\frac{r}{c}} \frac{dxdydz}{r} &=&\frac{c^2}{r_0}\; ,\nonumber\\
\frac{2m}{\hbar^2} A^{2}\int \frac{dxdydz}{r} &=&\frac{1}{r_0}\; ,\nonumber\\
 \frac{2mA^2}{\hbar^2}\frac{V}{r_0} &=& \frac{1}{r_0}\; ,\nonumber
\end{eqnarray}
i.e.
\begin{equation}\label{norm-Ros} 
  \frac{2mA^2V}{\hbar^2}=1 \; .
\end{equation} 
This condition is consistent from the point of view of dimensional analysis. To understand it, let us consider action (\ref{az-glob2}). The presence of the four-dimensional Einstein coupling $ \kappa $ produces a consequence for the length dimensions of the wave function $ \Psi $. We remember that the curvature scalar has dimensions $ \left[ \tilde{R}\right] =(length)^{-2} $ for every space-time dimension and we observe that from  (\ref{az-glob2}) it follows that $ \kappa\mathcal{L} $ and $ \tilde{R} $ have the same dimensions. As a consequence, the squared wave function amplitude $ A^{2} $ has the following dimensions $ [A^{2}]=\frac{(length)(mass)}{(time)^{2}} $ as it should, because of equation (\ref{norm-Ros}). It is worth noting that from Rosenfel's point of view, the wave function of a particle is not a point singularity: its amplitude is non zero in a finite volume $ V $. This fact is in contrast with de Broglie's point of view as Rosenfeld anticipated in the introduction of his paper. 

In this paper, Rosenfeld did not make any particular comment on (\ref{norm-Ros}) and on the whole calculation: he would discuss the physical meaning of the whole apparatus in the next papers, that we will briefly analyse in the following section. However, for us, Rosenfeld's calculation acquired a fundamental importance. Indeed, with this derivation the author showed for the first time how in the semi-classical limit GRQM is able to reproduce the RN metric in the weak-field approximation. In particular the condition (\ref{norm-Ros}) found by Rosenfeld can be interpreted as the normalization condition for the wave function. In this pre-second-quantized picture, the normalization condition of the wave function can be imposed using the definition of the Hamiltonian\footnote{In the weak-field limit, at the first order, the metric is flat.} (\cite{Landau-teo-quant-rel}) $ H $:
\begin{equation}
H = \int d^3x T_{00}\, ,
\end{equation}
where $ T_{00} $ is the $ 00 $ component of the total stress-energy tensor (\ref{tens-psi1}). The integration is carried out over the three-spatial volume for the following reason. The stress-energy tensor defined by Rosenfeld is a four-dimensional object, because of the unusual coupling between matter and geometry in the action (\ref{az-glob2}). The presence of the four-dimensional constant $ \kappa $ means that the stress-energy tensor's components represent an energy density with respect to the three-dimensional volume, instead of a four-dimensional volume. Rosenfeld was aware of this peculiarity, even if he did make no specific comment, because he noted that equations (\ref{eq-glob-5D}) imply a relation for the four-dimensional curvature scalar\footnote{See appendix \ref{app3d} for a detailed explanation.}, namely
\begin{equation}\label{def-dens-mass}
R=-\kappa\left[ \gamma^{\nu\bar{\mu}}T_{\bar{\mu}\nu} -\gamma^{\mu\rho}A_\rho\nabla_\lambda\left( \gamma_{\mu\sigma}F^{\sigma\lambda}\right) \right]\; , 
\end{equation}
that permitted him to define a four-dimensional mass density\footnote{Remember that in GR the trace of the stress-energy tensor is proportional to the curvature scalar and it is the energy density at first order in $ v/c $.} (\cite{Ros1}; p. 318, equation (63)), i.e. the quantity between the squared brackets on the r.h.s. of (\ref{def-dens-mass}). For a stationary charge, in the weak field limit, the four-dimensional density mass defined by Rosenfeld in (\ref{def-dens-mass}) coincides with $ T_{00} $. Moreover, for a localized wave packet the Hamiltonian must correspond to the rest energy $\mathcal{E}=mc^2 $ of the classical particle. In the case of a constant amplitude, the $ T_{00} $ value can be easily read off using equations (\ref{lagr-Ros}), (\ref{fun-onda}), (\ref{tens-psi1}), and the normalization condition for the wave function reads:
\begin{equation}\label{norma}
\int d^3x \frac{2m^2c^2A^2}{\hbar^2} = mc^2\quad\Rightarrow\quad  \frac{2m^2c^2A^2}{\hbar^2} V=mc^2 \quad\Rightarrow\quad \frac{2mA^2V}{\hbar^2}=1 \, ,
\end{equation}
where $ V $ is the three-volume of the localized wave packet. The normalization condition is precisely Rosenfeld condition (\ref{norm-Ros}). This normalization condition can be obtained also by considering the conserved current $ j^{\bar{\mu}} $. In the weak field approximation the continuity equation is $ \partial_{\bar{\mu}}j^{\bar{\mu}} = 0 $. Using the wave function Ansatz (\ref{fun-onda}) with a real constant amplitude $ A $, namely $ \Psi = A\, exp\left[ \frac{i}{\hbar}\left( -\frac{e}{c\beta}x^5+mcx^0\right) \right]  $, the continuity equation reads $ \displaystyle \frac{\hbar}{i}\frac{\partial\rho}{\partial t} = 0 $, where the squared modulus of the ``probability amplitude'' $ \rho $ is $ \rho = \frac{2m}{\hbar^{2}}A^2 $. By integrating over a three-spatial volume, because of the unusual length dimensions of the scalar field $  \Psi $, the normalization condition reads $ \frac{2mA^2V}{\hbar^2}=1 $, that is the same result obtained using the stress-energy tensor.

In the rest of his first paper, Rosenfeld tried to generalize his previous results to the case of a many-body system. This generalization process would continue in his following papers, where the author also analysed the role of the wave function amplitude $ A $. Rosenfeld inspected the consequences produced by considering a non-constant amplitude. In particular, he would be interested in its interpretation as a `potential of the internal forces' (\cite{Ros1}; p. 325) that should emerge when considering a continuous system. This idea was also shared by de Broglie, but was introduced by De Donder\footnote{The original citations are not quoted.}, as Rosenfeld wrote: `Recently, Mr. De Donder has introduced in WM two important concepts: the notion of \textit{permanence} of a system and the interpretation of the amplitude $ A $ of the Schr\"odinger's function $ \Psi $ as a \textit{potential of the internal tensions} of the system.'\footnote{We will not deepen the concept of ``permanence''.} (\cite{Ros2}; p. 447).

\subsection{The role of the correspondence principle in QG}\label{corr-princ}
As noted in our previous section, the first communication was sent to De Donder, who asked Rosenfeld to work with him during the summer of 1927. Even if they did not publish a joint paper, they cited each other in the communications published by the \textit{Bulletin de l'Acad\'emie royale de Belgique} \cite{DD-Ros}, \cite{Ros2}, \cite{Ros3}. Rosenfeld acknowledged De Donder explicitly at the end of the introduction: `My warmest thanks to Mr. De Donder, who did not quit to take an active interest in my work.' (\cite{Ros2}; p. 448). At the end of the third paper's introduction, Rosenfeld underscored again: `Mr. De Donder played an essential role in this work, because he suggested to me the basic idea. I owe a lot to De Broglie, who kindly continued to have a correspondence with me of which I took greatest advantage.' (\cite{Ros3}; p. 574). The main result of Rosenfeld-De Donder collaboration was the introduction of Bohr's correspondence principle as a physical interpretation of Rosenfeld's previous mathematical treatment. As far as we know, this is the first time that Bohr's principle was invoked in searching for a theory that could reconcile WM with GR. In particular, Rosenfeld and De Donder posed this principle as one of the founding principles of this new theory, which De Donder called `the gravitational wave mechanics' (\cite{DD-Ros}; p. 506). The purpose of this paragraph is to discuss the role of the correspondence principle, presenting Rosenfeld's following works: \cite{Ros2}, \cite{Ros3} and \cite{Ros5}. 

In order to understand the role of the correspondence principle, we start pointing out that Rosenfeld was impressed by the fact that the stress-energy tensor (\ref{tens-psi1}) resembled the stress-energy tensor for a particles' system whose form was:
\begin{equation}\label{tens-part}
T_{\mu\nu}=\sigma_{(m)}g_{\mu\rho}g_{\nu\sigma}u^{\rho}u^\sigma +P_{\mu\nu}\; , \qquad\text{where}\qquad u^\rho = \frac{dx^\rho}{d\tau}\; ,
\end{equation}
as it appears in De Donder's MIT lectures (\cite{DeDonder2} p. 52), and where $ \sigma_{(m)} $ represents the mass density as measured by the observer $ u^{\mu} $. For a swarm of non-interacting particles $ P_{\mu\nu} = 0 $, for a perfect fluid with pressure $ p $, $ P_{\mu\nu} = p\left( u_\mu u_\nu + g_{\mu\nu} \right) $ (\cite{MTW}); p. 132), while if we consider the dissipative processes its form is more complicated. The resemblance between the stress-energy tensor of a scalar field and that of a particle's system emerges as follows. Rosenfeld considered the following Ansatz for the wave function and for the Jacobi function:
\begin{eqnarray} 
\Psi \left( x \right)  &=& A\left( x^0\, ,x^1\, , x^2\, , x^3 \right) e^{\frac{i}{\hbar}\bar{S}} \label{fun-onda-gen}\\
\bar{S}\left( x\right) &=& -\frac{e}{\beta c} x^5+S\left( x^0\, ,x^1\, , x^2\, , x^3 \right) \; , \label{form-Jacobi}
\end{eqnarray}
where now $ S $ has an unspecified form and $ A $ is an arbitrary real function. The author inserted (\ref{fun-onda-gen}) into equation (\ref{tens-psi1}), and the stress-energy tensor components read:
\begin{equation}\label{emt}
T_{\bar{\mu}\nu}=2\frac{A^2}{\hbar^2}\partial_{\bar{\mu}}\bar{S}\partial_{\nu}\bar{S}+
2\partial_{\bar{\mu}}A\partial_{\nu}A + \gamma_{\bar{\mu}\nu}\mathcal{L}\, ,
\end{equation}
where the $ 55 $ component has been explicitly omitted, because Rosenfeld was not interested in the $ 55 $ component of five-dimensional Einstein equations.
Using the inverse components of the metric, equation (\ref{inverse}), Rosenfeld rewrote equation (\ref{RosB}), that we rewrite here for convenience
\begin{equation}
\gamma^{\mu\bar{\nu}}\partial_{\bar{\nu}}\bar{S}=mc\frac{dx^{\mu}}{d\tau}\; , 
\end{equation}
in the following form:
\begin{equation}\label{RosB2}
g^{\mu\nu}\partial_\nu S = mcu^\mu + \frac{e}{c}A^\mu\; .
\end{equation}
Equations (\ref{RosB2}) and (\ref{form-Jacobi}) imply that:
\begin{eqnarray}
\partial_{\mu}\bar{S} &=& \partial_\mu S = g_{\mu\nu} mcu^\nu + \frac{e}{c}A_\mu\; ,\label{S1}\\
\partial_5\bar{S} &=& -\frac{e}{\beta c}\; .\label{S2}
\end{eqnarray}  
Inserting equations (\ref{S1}) and (\ref{S2}) in (\ref{emt}), the author obtained\footnote{Equation (\ref{emt-em}) was obtained raising an index with the five-dimensional metric, $ \displaystyle \gamma^{\bar{\rho}\bar{\mu}}T_{\bar{\mu}\bar{\nu}} $, and then choosing $ \bar{\rho}=\rho $ and $ \bar{\nu}=5 $.} (\cite{Ros2}; p. 454):
\begin{eqnarray}
T_{\mu\nu} &=&\varrho_{(m)} g_{\mu\rho}g_{\nu\sigma}u^{\rho}u^\sigma + \Pi_{\mu\nu}\label{emt-spaz}\\
\beta {T_5}^{\nu} &=& \varrho_{(e)} u^\nu + \Lambda^\nu \label{emt-em}\, ,
\end{eqnarray}
where we define, following Rosenfeld, a ``quantum'' mass density $ \varrho_{(m)}  $ and a ``quantum'' charge density\footnote{Remember that ab inizio we decided to consider the case of $ q=-e $.} $ \varrho_{(e)} $:
\begin{equation}\label{density}
\varrho_{(m)}= \frac{2m^2c^2}{\hbar^2}A^2\qquad\quad\qquad\varrho_{(e)}=-\frac{2em}{\hbar^2}A^2\, .
\end{equation}
Equations (\ref{emt-spaz}) and (\ref{emt-em}) require some comments, because, from Rosenfeld's and De Donder's point of view they are the basis for invoking the correspondence principle.

Firstly, the analogy between (\ref{tens-part}) and (\ref{emt-spaz}) is now evident, and this explains why $ \varrho_{(m)} $ could play the role of a mass density. In order to understand why $ \varrho_{(e)} $ represents a charge density, we remember that the Maxwell equations on curved space-time for a classical charged system are
\begin{equation}\label{def-e}
\nabla_\mu F^{\nu\mu} = j^\mu\qquad \text{whith} \qquad j^\mu = \sigma_{(e)}u^\mu\; ,
\end{equation}
where $ \sigma_{(e)} $ represents the charge density of the system as measured by the observer $ u^{\mu} $. On the other hand, the Maxwell equations obtained by the five-dimensional Einstein equations coupled to the wave function stress-energy tensor (\ref{eq-glob-5D}) are\footnote{See the appendix \ref{app3c} for technical details.}:
\begin{equation}\label{maxwell}
\nabla_\mu F^{\nu\mu}=\beta{T_5}^\nu\, .
\end{equation}
Therefore, it is evident that $ \beta {T_5}^{\nu}  $ could play the role of the density current $ j^{\mu} $ and, as a consequence, equation (\ref{emt-em}) defines a charge density $ \varrho_{(e)} $. 

Secondly, this is the point where Bohr's principle comes into play. At the end of the introduction of his communication, Rosenfeld underscored that the identification of $\varrho_{(m)}$ and  $ \varrho_{(e)} $ with the mass and electric densities of quantum system is `a particularly instructive aspect of the \textit{correspondence principle}' (\cite{Ros2}; p. 448): he stressed that this claim would deserve further analysis and that the connection between the above identification and the correspondence principle has been suggested by De Donder. At the end of the fifth section of the brief communication, Rosenfeld remarked that (we changed the original equation's numbers in order to fit our numerical order): `equations (\ref{emt-spaz}) and (\ref{emt-em}) show that  $\varrho_{(m)}$ and  $ \varrho_{(e)} $ should be interpreted as a mass density and an electric density of the system, or, better$ (^*) $, \textit{corresponding} to the system [...]' (\cite{Ros2}; p. 454). Rosenfeld himself used the italics and in the footnote corresponding to the symbol $ (^*) $ he underscored again that this remark had been suggested by De Donder. The term ``corresponding'' referred to the formal correspondence between a classical and a quantum system. Indeed, $ \varrho_{(m)} $ and $ \varrho_{(e)} $ depend on the wave function's amplitude. In the following papers, Rosenfeld would clarify how his approach is connected with Bohr's correspondence principle. Our last comment concerns the terms $ \Pi_{\mu\nu} $ and $ \Lambda_\nu $. Their precise form will not be discussed here, but it is worth noting that they contain the contribution due to the fact that the amplitude is not constant. From Rosenfeld's and De Donder's point of view the $ \Pi_{\mu\nu} $ tensor would represent the contribution of the internal forces of the system, while $ \Lambda_\nu $ was called `quantum current' (\cite{Ros5}; p. 665) by Rosenfeld, maybe because it has no classical analogue. 

In the third communication Rosenfeld dedicated an entire section to enunciate his principle of correspondence, explicitly referring to Bohr's principle, also describing what he had in mind as QG theory (we changed the original equation numbers in order to fit our numerical order): 
\begin{quote}
`The wave mechanics obtained using the variational principle (\ref{az-glob2}) realizes \textit{formally} the fusion between Gravity and quantum theory. To the \textit{field equations} that describe gravitational and electromagnetic phenomena, we added the \textit{equation of quantization} (\ref{KG4}), that rules the quantum-energy exchanges. In this last equation intervenes the \textit{fundamental quantity} $ \Psi $, and the fusion between the two theories is represented by the fact that the five-dimensional matter tensor that is present in the [gravitational] field equation is defined using the fundamental quantity $ \Psi $; on the contrary, in a \textit{pure} Einsteinian gravitational theory, this tensor is a function of different fundamental quantities of the system: the \textit{mass density} $ \sigma_{(m)} $ and the \textit{electric charge density} $ \sigma_{(e)} $.'\footnote{The term `\textit{pure} Einsteinian gravitational theory'  seems to be referred to the classical theory obtained without the introduction of the ``quantum field''. We introduced Rosenfeld symbols $ \sigma_{(m)} $ and $ \sigma_{(e)} $ in equations (\ref{tens-part}) and (\ref{def-e}) respectively.\label{pure}} (\cite{Ros3}; p. 574). 
\end{quote}
Rosenfeld used different letters referring to the mass and charge densities, because he wanted to emphasize the difference between a classical system and the corresponding quantum system. The author continued: 
\begin{quote}
`The new definition of the stress-energy tensor as a function of $ \Psi $, (\ref{tens-psi1}), implies a modification of our conception for the role of the fundamental quantities  $ \sigma_{(m)} $ and $ \sigma_{(e)} $. In the Einsteinian theory these quantities intervene directly in in the field equations in order to fix the gravitational and the electromagnetic potentials, corresponding to a given distribution $\left(  \sigma_{(m)}\, ,\,  \sigma_{(e)}\right)  $. In Wave Mechanics, these quantities do not intervene directly, but through [..] the quantity $ \Psi $. [...] The material tensor $ T^{\bar{\mu}\bar{\nu}} $ as a function of $ \Psi $ should not necessarily be identical to the material tensor of \textit{pure} Gravity, which is defined as a function of $ \sigma_{(m)} $ and $ \sigma_{(e)} $. It seems desirable to analyse, thenceforward, as soon as possible, the behaviour of the $ T^{\bar{\mu}\bar{\nu}} $ tensor, \textbf{in order to emphasize all possible modifications to Gravity produced by the introduction of the quantum quantity} $ \Psi $; this is the role of the \textit{principle of correspondence}. [bold form added]'\footnote{The term \textit{pure Gravity} can be interpreted as GR. See also footnote (\footref{pure}).} (\cite{Ros3}; p. 575). 
\end{quote}
The bold text emphasizes clearly what was the physical meaning of the calculation presented in section \ref{calculation}. From Rosenfeld's point of view, the introduction of the wave function was responsible for the modifications of the ``pure'', i.e. classical, GR, because even in the case of constant amplitude, it permits us to introduce two quantum quantities, corresponding to classical quantities $ \sigma_{(m)} $ and $ \sigma_{(e)} $: through the new stress-energy tensor, the new quantities $ \varrho_{(m)} $ and $ \varrho_{(e)} $, defined by (\ref{density}), must be considered as the quantum source of gravitational and electromagnetic field. Indeed Rosenfeld continued: 
\begin{quote}
`The comparison between $ \varrho_{(m)} $ and $ \varrho_{(e)} $, and $ \sigma_{(m)} $ and $ \sigma_{(e)} $ will show us how the quantum objects will modify the gravitational and the electromagnetic phenomena. It will be possible to enunciate a more precise and \textit{general} correspondence principle; [...] there are some precise formulas that define, in a strict sense, the principle of correspondence and that establish the identification of the formal schema of wave mechanics with the gravitational schema of Th. De Donder, [...] showing how Wave Mechanics widens the picture of the pure Gravity, in order to incorporate quantum phenomena.' (\cite{Ros3}; p. 575). 
\end{quote}
It is important to stress that, like Klein, de Broglie and De Donder, Rosenfeld never discussed the role of the boundary conditions of the wave function. Like De Donder he referred to the introduction of the wave function as the `equation of quantization'. It is worth to remember that Heisenberg's uncertainty principle was introduced in February of the same year \cite{Heisenberg}. This coincided with the fact that Rosenfeld considered it sufficient to introduce the wave function into Einstein's equations in order to describe correctly the coupling between gravity and quantum matter. 

Rosenfeld did not cite any of Bohr's papers, but the idea that the correspondence principle could be a theoretical argument to infer the behaviour of a quantum system with respect to the classical one is a consequence of Bohr's influence. Indeed, in the introduction of the third communication, Rosenfeld declares that his approach, i.e. the variational principle, is a `formal theory'  (\cite{Ros3}; p. 573). Then he continued: `To put a physical interpretation [on the formal theory], we let ourselves be guided by the \textit{correspondence principle}, using the interpretation given by O. Klein \cite{Klein6} ...' (\cite{Ros3}; p. 573). 

In order to understand Bohr's role, we briefly analyse Klein's paper \cite{Klein6}. Klein's work is a cornerstone of the history of QM. Before that article, matrix mechanics was the only approach incorporating the correspondence principle\footnote{Heisenberg referred to the fact that classical results can be obtained, in matrix mechanics approach, in the limit of high quantum numbers.}, as Heisenberg himself reported in his review of matrix mechanics' successes in 1926 (\cite{history-QM}; p. xxxii). In this sense, the title of Klein's contribution was very revealing: \textit{Electrodynamics and Wave Mechanics from the point of view of the Correspondence Principle}. As reported in \cite{history-QM}, Bohr was aware of the content of Klein's work and he expressed an enthusiastic comment in a letter to Schr\"odinger (\cite{history-QM}; p. 176). In particular, Bohr was fascinated by the connection between Hamiltonian mechanics and HJ dynamics of wave rays, that generated Klein's relativistic WM. Paraphrasing Bohr's words, he was interested in the fact that thanks to this analogy it is possible, on the basis of WM, to build a corresponding theory. Klein's main purpose was to investigate the possibilities of exploiting relativistic WM for understanding atomic processes involving discontinuities. In Klein's paper, the correspondence principle intervenes when the author tries to modify Maxwell's equations. Schr\"odinger also expressed the idea that the wave function `possesses the property to enter even the untouched [classical] Maxwell-Lorentz equations between the electromagnetic field vectors as a ``source'' of the latter' (\cite{history-QM}; p. 43). 

In 1927 Schr\"odinger investigated also the effect on the stress-energy tensor obtained by a unified variational principle involving the Maxwell's Lagrangian and the complex scalar field Lagrangian, i.e. `the de Broglie's wave' (\cite{Schroe2}; p. 265). Unlike Klein, de Broglie and Rosenfeld, Schr\"odinger declared explicitly that he would consider neither additional dimensions, nor gravitational field contributions. Indeed, Schr\"odinger's Lagrangian $ \mathcal{L}_S $ is the sum of Maxwell's Lagrangian, $\displaystyle \mathcal{L}_{em} =-\frac{1}{4}F_{\mu}F^{\mu\nu} $, and $ \mathcal{L}_\psi $, the Lagrangian for material fields, which is related to De Donder's work, see (\ref{lagr-DD}), because Schr\"odinger cited De Donder's contribution:
\begin{equation}
\mathcal{L}_S =\mathcal{L}_{em} + \mathcal{L}_\psi =-\frac{1}{4}F_{\mu}F^{\mu\nu} -\eta^{\mu\nu}\left( \partial_{\mu}\psi+\frac{i}{\hbar}\frac{e}{c}A_\mu\psi\right) \left( \partial_{\nu}\overline{\psi}-\frac{i}{\hbar}\frac{e}{c}A_\nu\overline{\psi}\right)+\frac{m^2c^2}{\hbar^2}\overline{\psi}\psi \; .
\end{equation}
$ \mathcal{L}_S $ can be obtained after a dimensional reduction from Rosenfeld's Lagrangian (\ref{az-glob2}) in the limit of a flat background. But  Schr\"odinger did not investigate the role of $ \psi $ as a source of the electromagnetic field, because he explicitly asserted that the KG Lagrangian $ \mathcal{L}_\psi $ did not describe any real field. In spite of this, Klein analysed this aspect, inspired by the idea to use the correspondence principle. First he manipulated his scalar relativistic equation to define the four-vector $ j^\mu = \left( \rho\, ;\, j^i \right) $, where\footnote{The symbols have the usual meaning. We remember that the electromagnetic potentials are $ A^\mu = (A^0 ; A^i) $} (\cite{Klein6}; p. 414, equations (20)):
\begin{eqnarray}
\rho &=& -\frac{e}{2mc^2}\left\lbrace -\frac{\hbar}{i}\left( \overline{\psi}\frac{\partial\psi}{\partial t}-\psi\frac{\partial\overline{\psi}}{\partial t}\right)+2e\overline{\psi}\psi A^0 \right\rbrace \label{j1}\\
j^i &=& -\frac{e}{2m}\left\lbrace \frac{\hbar}{i}\eta^{ij}\left( \overline{\psi}\partial_i\psi-\psi\partial_i\overline{\psi}\right)+2\frac{e}{c}\overline{\psi}\psi A^i \right\rbrace \; . \label{j2}
\end{eqnarray}       
Then he showed that using the usual optical geometric Ansatz $ \psi=e^{\frac{i}{\hbar}S} $ for the wave function, in the semiclassical limit $ \hbar\rightarrow 0 $, equations (\ref{j1}) and (\ref{j2}) reduce to the components of the usual potentials for a relativistic scalar charged particle, namely:
\begin{eqnarray}
\rho_{cl} &=& -\frac{e}{\sqrt{1-\left( v^2 / c^2 \right) }} \label{j1a}\\
j^i_{cl} &=&  -\frac{ev^i}{\sqrt{1-\left( v^2 / c^2 \right) }}\; , \label{j2a}
\end{eqnarray}
where $ v^i $ is the three-velocity of the particle\footnote{The role of the analogy between Hamiltonian dynamics and the dynamics of wave's rays is fundamental to obtain these relations.} and $ v $ its modulus. Finally, using the correspondence principle, Klein interpreted equations (\ref{j1}) and (\ref{j2}) as the source for the electromagnetic field, in order to investigate the quantum modifications of the Maxwell equations, namely:
\begin{eqnarray}
\partial_i E^i &=& 4\pi\rho \label{j1b}\\
\varepsilon^{ijk}\partial_j B_k - \frac{1}{c}\frac{\partial E^i}{\partial t} &=& \frac{4\pi}{c}j^i\; . \label{j2b}
\end{eqnarray}
Klein solved the Maxwell equations (\ref{j1b}) and (\ref{j2b}) using the advanced and the retarded potentials, in order to write an expression for the electric and the magnetic fields as functions of $ \psi $. Klein identified these electric and magnetic fields with the electromagnetic field produced by the bounded electron\footnote{Unlike Rosenfeld, Klein considered also the full quantum treatment, introducing the eigenfunctions expansion for the wave field.}, by means  of the correspondence principle (\cite{Klein6}; p. 422, equations (41). See also equations (33), (28) and (18)). As we have seen, Rosenfeld followed the same path in order to obtain an expression for the metric components, explicitly referring to Klein's paper. In this sense, Rosenfeld was the first author to introduce the correspondence principle in the context of QG. It is worth noting that in the five-dimensional picture the Maxwell equations are naturally coupled to the four-current, like Rosenfeld himself showed with relations (\ref{maxwell}). This seemed to be another advantage of the five-dimensional approach.

\subsection{Back to the present}\label{discussion}
In his last paper of the year\footnote{In a brief communication to the \textit{Comptes rendus} in June of the same year, Rosenfeld claimed that he was able to reproduce Epstein's description of `the magnetic electron of Uhlenbeck and Goudsmith' (\cite{Ros4}; p. 1541), i.e. the spinning electron, using the five-dimensional apparatus described in the previous section. We will not go into the reasons that could explain Rosenfeld's claim, because we postpone this analysis to a future work.}, written in October 1927, Rosenfeld made a detailed and wider exposition of all the concepts introduced in his previous work. His idea was to formulate a sort of formal basis for the five-dimensional Universe as a unified framework for GR and WM. The foundations of the whole building are three principles: a variational principle, i.e. equation (\ref{az-glob2}); the principle of Schr\"odinger eigenfunctions, i.e. the usual `boundary conditions that must be imposed on $ \Psi $ and $ \overline{\Psi} $ in order to quantize the system' (\cite{Ros5}; p. 665); and the correspondence principle, that the author formulated with the help of De Donder. Rosenfeld also cited a paper written by De Donder, where the latter tried to give a more precise formulation of the principle \cite{DD-Ros}. Unlike Rosenfeld, De Donder will not abandon this idea in the future. Indeed while Rosenfeld seemed to be convinced that quantum theory should modify GR, De Donder will continue to claim that GR and WM, were compatible theories \cite{DD-1930}.

Rosenfeld confirmed the ideas proposed in the previous paper, claiming that the components of the new stress-energy tensor as a function of the wave function $ \Psi $ should play the role of `quantum currents', i.e. quantum source for the right side of Maxwell and Einstein equations. The author wrote explicitly: `The \textit{correspondence principle} consists in stating that this analogy is not only a formal analogy, but also a physical analogy.' (\cite{Ros5}; p. 666). He also emphasized the particular nature of the correspondence principle: `There exist \textit{postulates} in the sense of the formal logic, whilst the correspondence principle is a \textit{physical principle} [...]' (\cite{Ros5}; p. 667). Rosenfeld meant that the extension of the analogy from the formal plane to the physical plane is a sort of meta-sentence, and it was different, in this sense, from a formal sentence of the ``basic language'' of the equations, like e.g. the variational principle.

Rosenfeld's approach, as well as de Broglie's proposal were briefly discussed at the Solvay conference. As stated above, in section 2, Rosenfeld was not officially admitted to the conference, but De Donder invited him to follow him, in order to have the possibility to meet Pauli at the conference. The conference's proceedings showed once again how de Broglie, Rosenfeld and De Donder agreed on the meaning of the five-dimensional Universe. De Broglie asserted that De Donder  succeeded in harmonizing Einstein theory with WM (\cite{Crossroad}; p. 483); De Donder tried to draw attention to  the MIT lectures we previously discussed, speculating on a connection between his correspondence principle and Bohr reflections (\cite{Crossroad}; p. 483). Subsequently De Donder stated that there is a connection between de Broglie's contributions, his work and Rosenfeld's ideas (\cite{Crossroad}; p. 499 and 519). De Donder will try again to discuss his approach (\cite{Crossroad}; p. 470; 471; 510), but the questions raised by De Donder and de Broglie will not be faced by the group of physicists.

De Donder's approach to Hamiltonian dynamics discussed in section 2 is peculiar, because he introduced systematically the use of poly-momenta $ p_\mu $ obtained starting with a Lagrangian $ \mathcal{L}(y^a\, ,\, \partial_\mu y^a ) $, which were functions of some variables $ y^a $ and its derivatives, deriving it with respect to all of the derivatives, $p^a_\mu = \frac{\partial\mathcal{L}}{\partial\partial_\mu y^a} $, instead of using the time derivative only as usual. This convention, sometimes called the De Donder-Weyl approach, and its generalization to a curved space-time has survived until recent years, as an alternative approach for the quantization of gravity, and it is today known as pre-canonical quantization \cite{Kanatchikov1} \cite{Kanatchikov2}.

At the end of 1929, after his stay in G\"ottingen, Rosenfeld moved to Z\"urich where, stimulated by Pauli, tried to inspect what we today call the gravitational self-energy of a quantized electromagnetic field. In \cite{Rosenfeld1} he approached the problem in a way that resembles the work analysed here. Like in his previous work, he integrated again the linearised Einstein equations, this time coupled with Maxwell equations only. The quantized electromagnetic field played the role of the complex scalar field. Rosenfeld used the annihilation and creation operators approach for treating the electromagnetic field, hence the metric field $ h_{\mu\nu} $ itself was described by an operator. In this sense he obtained again a sort of quantum metric, because it is generated by a quantum field. Rosenfeld did not cite the previous papers we analysed, but we must stress the importance played by his early work, because of the affinity of the path followed by the author. 

The term quantum metric could be understood in a complementary way. The quantum corrections to the classical gravitational field can be considered as the contribution to the classical effects produced by the quantization of the gravitational field. In the mid Thirties, Matvei Bronstein \cite{Bronstein1} would quantize for the first time the gravitational field directly in the weak field limit, in order to understand quantum deviations from the classical Newton law. Only 37 years later, after the development of perturbation theory, Micheal Duff \cite{Duff-72} tried to understand the quantum corrections to the Schwarzschild metric. Duff used explicitly a classical source and he quantized directly the gravitational field. At the tree level, in the weak field limit, he obtained the classical results, while the quantum corrections came from the one-loop corrections.

Finally we address the following question: what is the physical meaning of Rosenfeld's result from the modern point of view? Rosenfeld interpreted the particle's wave function as the source of the gravitational field. From a modern point of view, this approach treats the gravitational interaction as a classical phenomenon and the particle's description as fully quantized. This means that Rosenfeld's procedure gives a semi-classical result, even in the case of non constant amplitude. From a modern point of view, Rosenfeld's results can be obtained as non-relativistic limit of the so-called semi-classical Einstein equations, an approach formally suggested by M\o{}ller for the first time \cite{Moller}. These equations are obtained by replacing the stress-energy tensor, i.e. the r.h.s. of Einstein equations, by the expectation value of the stress-energy operator $ \hat{T}_{\mu\nu} $ with respect to some quantum state $ \ket{\Psi} $. In four dimensions they have the following form:
\begin{equation}\label{sEe}
R_{\mu\nu}-\frac{1}{2}g_{\mu\nu}R = 8\pi G \bra{\Psi}\hat{T}_{\mu\nu}\ket{\Psi}\, .
\end{equation}
The modern interpretation of equation (\ref{sEe}) is connected with the character of the coupling between gravity and matter. This character has not yet been clarified and it is an open problem in the QG research area. It is equivalent to the question whether gravity should be quantized or not \cite{Treder}. This is a long debate, see e.g. \cite{Carlip} and \cite{Kiefer}, that divided the physicist community in two groups, initiated incidentally by Rosenfeld himself \cite{Rosenfeld63}. On one side those who believe that the gravitational interaction must be quantized, on the other side those who believe that gravitational interaction must remain classical. As a consequence, for the first group equations (\ref{sEe}) can be derived approximately from canonical QG as a kind of mean-field equation \cite{Kiefer}. In this case, the metric obtained integrating the linearised Einstein equations following Rosenfeld's procedure is a sort of ``mean metric'' $ \bra{\Psi}\hat{g}_{\mu\nu}\ket{\Psi} $, where the hat-symbol means that the metric should be an operator. This perspective is also shared by those who investigate the behaviour of QFT on a curved background \cite{Birrel-Davies}, that led to Hawking's results on black hole's entropy. For the second group the coupling between quantum fields and classical gravity described by Einstein equations should be understood as a fundamental description of nature. As a consequence, they interpret the l.h.s. of (\ref{sEe}) as evaluated using the classical metric. From this perspective, a possible starting point for reconciling WM with gravity is the so called Schr\"odinger-Newton equation\footnote{The Schr\"odinger-Newton equation was introduced by Roger Penrose to provide a dynamical description of the quantum wave function's collapse \cite{Penrose}.} \cite{Grossart}, where the source of the gravitational field is represented by the squared modulus of the wave function. We do not enter the debate whether which approach could be the fundamental one, because we believe that any extension of our conceptual framework for the description of nature would be of interest in itself. We observe that recently there has been a revival of Rosenfeld's ideas coming from the second group of physicists. Modern authors, \cite{Giulini} and \cite{Grossart}, with different scope, used some of the Rosenfeld's ideas, extended to the non-static case. More precisely, in \cite{Giulini} the authors studied the coupling between KG field and gravity in the case of a non-static spherical symmetric space-time, in the limit of semi-classical and non-relativistic approximation from the four-dimensional point of view. Following Kiefer's scheme for non-relativistic and semi-classical approximation, the authors investigated KG equation on a curved background, showing that it reduces itself, in this WKB-like scheme, to the Newton-Schr\"odinger equation, at a certain order of the WKB expansion. Einstein equations coupled with the KG stress-energy tensor reduces, in the same approximation, to the Poisson equation for the gravitational potential, where the wave function amplitude plays the role of the mass density. This means that, like in Rosenfeld's scheme, the wave function is the source of the metric. At the order chosen by the authors, the metric itself results as an expansion in terms of $ \frac{\hbar}{c^2} $ powers and it depends on the wave amplitude of the field. In the weak field limit, the quantum-mechanical description can be derived from the field-theoretic approach with a well defined procedure, which allows one to use the wave function, instead of Fock's states \cite{QFT-QM}. In \cite{Grossart}, the authors refined their analysis using the second-quantised formalism and hence they apply the procedure to find the quantum mechanical limit. Once again they find that the wave function is the source of the gravitational field, like in Rosenfeld's approach.

\section{Summary and Conclusions}\label{summary}
In this paper we have described the earliest of Rosenfeld's contributions of 1927. From an historical point of view, Rosenfeld's work is interesting for various reasons. First, it contains many ingredients that the author will use in his future work. Second, it shows how Rosenfeld was influenced by his mentors: Oskar Klein, Louis de Broglie and Theophile De Donder. Third, it offers a connection between the history of QM and the history of QG. 

We started considering the main results achieved by his mentors, at the time he started to write his first paper. Klein wrote a five-dimensional unified variational principle for the electromagnetic and the gravitational field. He introduced the relativistic wave equation on a curved background using the correspondence between Hamiltonian dynamics for point particles and the HJ equation in the geometrical optics limit. Following this correspondence, Klein tried to introduce a sort of massless KG equation, in analogy with light. De Broglie was pressed by Rosenfeld, who joined the French physicist in Paris, to investigate the five-dimensional Universe features. De Broglie showed that it is not necessary to consider null-geodesics, and that the four-dimensional geodesics can be represented as the projection of five-dimensional geodesics. De Broglie built his five-dimensional Universe using an inconsistent time-like extra dimension, as Klein himself would note in a following paper. De Donder, the third character of our story, introduced the Lagrangian approach involving the wave function, treating it as a field, again using the correspondence between Hamiltonian particle dynamics and the HJ equation for wave's rays. De Donder interpreted the introduction of a unified variational principle as the mathematical instrument responsible for the quantization of the system, because it produces the KG equation. He was convinced that no modifications of GR were needed for describing quantum phenomena. De Donder played a fundamental role in Rosenfeld's work. Rosenfeld sent De Donder his first paper, who presented it for publication at the \textit{Bulletin de l'Acad\'emie royale de Belgique} journal. Even though we have not analysed any De Donder-Rosenfeld correspondence, a collaboration between these authors emerges clearly. Furthermore, De Donder invited Rosenfeld to the fifth Solvay conference, where De Donder tried to draw attention to Rosenfeld's work and where Rosenfeld met Einstein and the physicists of the G\"ottingen school.     

After having introduced Klein's, de Broglie's and De Donder's approaches, we considered Rosenfeld's work. In his first paper, Rosenfeld tried to walk one step ahead with respect to his mentors. He decided to put De Donder's action model in a five-dimensional context, building upon the work of Klein and de Broglie. His second contribution, central in our analysis, was to address the task of understanding which metric can be generated by a quantum object, i.e. a localized electron's wave function. Rosenfeld tried also to understand which conditions must hold in order that WM and GR could reproduce in a semi-classical approximation a classical metric in the weak field limit. Studying this problem he presented for the first time a quantum modification of the flat metric, because of the appearance of $ \hbar $. In his following papers, thanks to De Donder's collaboration, Rosenfeld succeeded in giving a physical meaning to his mathematical treatment. De Donder recognized the idea of Bohr's correspondence principle in using the wave function's stress-energy tensor as a source of the gravitational field. In his third communication Rosenfeld himself explicitly recognized that his approach to QG was inspired to what Klein did in the context of Maxwell's equations.

Thanks to De Donder, Rosenfeld started to interact with Pauli, Jordan, Bohr himself and many other physicists who will play, unlike de Broglie and De Donder, a fundamental role in constructing the new quantum theory of fields. After 1927, Rosenfeld will convince himself of the importance of quantizing these new objects and, stimulated by Pauli, he will study again the problem of a quantum metric, but using the new-born quantum theory of massless spin-1 fields \cite{Rosenfeld1}. From an historical point of view, this paper concluded what we called the prehistory era in the history of QG.

Even if he never considered his early papers on QG an important work, Rosenfeld's contributions show how the search of a theory that could reconcile quantum phenomena with GR started early and that it also reached interesting results, that will continue to be valid in the context of quantum field theory on a curved space-time. Even if Klein, de Broglie, De Donder and Rosenfeld were not a research group as in our modern meaning, in 1927 their works were related by a common purpose.   

The problem of finding a quantum theory of gravity has never been limited, and is not limited today, to the quantization of gravitational interaction only. We now know that attempts to apply directly to the gravitational field quantization procedures, which have been successful in other contexts, have failed. From the beginning of the prehistory of QG, the authors that tried to face the problem of reconciling quantum phenomena with gravity interpreted the idea of QG in the broadest sense. From an historical point of view, the following statement is particularly true: `In the broadest sense, a quantum theory of gravitation would represent an extension of our conceptual framework for the description of nature: any such extension would be interest in itself.' (\cite{Ashtekar}; p.1213).

\section*{Acknowledgement}

We express our gratitude to all anonymous referees who gave us the opportunity of improving the original manuscript. We are very grateful to Kurt Lechner for his invaluable comments and suggestions.

\noindent{This work has been supported in part by the DOR 2016 funds of the University of Padua.}

\section{Appendices}\label{apps}

\subsection{Wave optics and null-geodesics in Klein's five-dimensional manifold}\label{geom-optic}
Klein's original idea was to write a wave equation in analogy with light in the context of his five-dimensional Universe. This appendix follows Klein's original approach \cite{Klein1}.

In a curved five dimensional space-time, a relativistic generalization of Schr\"odinger equation is represented by the following equation:
\begin{equation}\label{KG1bis} 
a^{\bar{\mu}\bar{\nu}}\left( \delta^{\bar\sigma}_{\bar\nu}\frac{\partial}{\partial x^{\bar{\mu}}} -\Gamma_{\bar{\mu}\bar{\nu}}^{\bar{\sigma}}\right)\partial_{\bar{\sigma}}\Psi= a^{\bar{\mu}\bar{\nu}}\nabla_{\bar{\mu}}\partial_{\bar{\nu}}\Psi=0\; ,
\end{equation}
where $ \Psi $ is the wave function and the covariant derivative $ \nabla_{\bar{\mu}} $ is defined using the Christoffel symbols $  \Gamma_{\bar{\mu}\bar{\nu}}^{\bar{\sigma}}$. As stated in the main text, Klein defined the Christoffel symbols using the space-time metric $ \gamma_{\bar{\mu}\bar{\nu}} $, that we rewrite here for convenience:
\begin{equation}\label{dsigma1}
d\sigma ^2  = \alpha d\theta ^2 + ds^2 \, , 
\end{equation}
\begin{center}
	where
\end{center}
\begin{equation}\label{dsigma2}
d\theta = dx^5 + \beta A_\mu dx^{\mu}\quad ;\quad 
g_{\mu\nu} = \gamma_{\mu\nu} - \frac{16\pi G}{c^4}A_{\mu}A_{\nu}\quad ;\quad ds^2=g_{\mu\nu}dx^{\mu}dx^{\nu}\; \, .  
\end{equation}
Equation (\ref{KG1bis}) resembles a massless equation for a scalar field, where the inverse of the space-time metric $ \gamma^{\bar{\mu}\bar{\nu}} $ has been replaced by the tensor $ a^{\bar{\mu}\bar{\nu}} $, whose covariant components are defined by equation (\ref{metric-K}). As stressed in sections \ref{section-Klein} and \ref{section-deBroglie}, this fact generated the ambiguity in Klein's approach, criticized by de Broglie. Following Klein's approach, we shall show how wave equation (\ref{KG1bis}) is connected with five-dimensional null-geodesics that reduce to the four-dimensional equations of motion for charged massive particles in a combined electromagnetic and gravitational field.

In the geometrical optics limit a wave front propagates locally as a plane-fronted wave. Therefore, the Ansatz for the wave function is
\begin{equation}\label{psi-app}
\Psi (x) = A e^{i\omega S(x)}
\end{equation}
where $ \omega $ is so large that only the term proportional to $ \omega^2 $ in equation  (\ref{KG1bis}) need to be taken into account. The function $ S=S(x) $ is termed the eikonal and it characterizes the phase of the wave. Substituting (\ref{psi-app}) into the wave equation, the term with two derivatives is proportional to $ \omega^2 $ and equation (\ref{KG1bis}) reads:
\begin{equation}\label{geom-optik}
a^{\bar{\mu}\bar{\nu}}\partial_{\bar{\mu}}S\partial_{\bar{\nu}}S = 0\,\,\, .
\end{equation}   
Last equation resembles the eikonal equation for light rays, that describes the propagation of the wave front $ S(x) $ of light rays. In the HJ approach, it can be derived by a particular Hamiltonian, whose Hamilton equations describe the dynamics of the particle associated to the wave by wave/particle duality. Klein we defined the Hamiltonian as follows:
\begin{equation}
H = \frac{1}{2}a^{\bar{\mu}\bar{\nu}}p_{\bar{\mu}}p_{\bar{\nu}}   \qquad\text{where}\qquad p_{\bar{\mu}} = \partial_{\bar{\mu}}S\;\,\, .
\end{equation}  
Hence, equation (\ref{geom-optik}) now reads:
\begin{equation}
H = 0  \; ,
\end{equation}
and parametrizing the five-dimensional particle's world line with an arbitrary parameter $ \hat{\lambda} $, the relativistic Hamilton equations are:
\begin{equation}
\frac{\partial H}{\partial p_{\bar{\mu}}}=\frac{dx^{\bar{\mu}}}{d\hat\lambda} \qquad ;\qquad
-\frac{\partial H}{\partial x^{\bar{\mu}}}=\frac{dp_{\bar{\mu}}}{d\hat\lambda}\; .
\end{equation}
The analogy between equation (\ref{geom-optik}) and the usual eikonal equation suggests to consider null-geodesics for the differential form $ a_{\bar{\mu}\bar{\nu}}dx^{\bar{\mu}}dx^{\bar{\nu}} $ as stated by Klein, where $ a_{\bar{\mu}\bar{\nu}} $ represent the reciprocal quantities of $ a^{\bar{\mu}\bar{\nu}} $. As emphasized in the main text, After a Legendre transformation, the Hamiltonian $ H $ is mapped into the following Lagrangian:
\begin{equation}
L= \frac{1}{2} a_{\bar{\mu}\bar{\nu}}\frac{dx^{\bar{\mu}}}{d\hat\lambda}\frac{dx^{\bar{\mu}}}{d\hat\lambda}\;\, ,
\end{equation}
where the covariant components of the tensor $ a_{\bar{\mu}\bar{\nu}} $ are:
\begin{equation}\label{metric-K1} 
a_{\mu\nu} = g_{\mu\nu}+\frac{e^2}{m^2c^4}A_\mu A_\nu \quad\quad a_{\mu 5}=\frac{e^{2}}{m^2c^3\beta}A_\mu \quad\quad a_{55}= \frac{e^2}{m^2c^4\beta^2}\;  \, .
\end{equation}
Like all the quantities introduced by Klein, also the components of $ a_{\bar{\mu}\bar{\nu}} $ do not depend on the fifth coordinate. As we emphasized in the main text, $ a_{\bar{\mu}\bar{\nu}} $ and $ \gamma_{\bar{\mu}\bar{\nu}} $ are quite different, cfr. equations (\ref{metric-K1}) and equations (\ref{dsigma2}). As we said, it seems that Klein introduced a new metric for the microscopic world, $ a_{\bar{\mu}\bar{\nu}} $, indeed the null-like character of the paths is referred to the tensor $ a_{\bar{\mu}\bar{\nu}} $ instead of $ \gamma_{\bar{\mu}\bar{\nu}} $. If following Klein we define $ \mu=a_{55} $, hence $ a_{\bar{\mu}\bar{\nu}}dx^{\hat{\mu}}dx^{\hat{\nu}} = \mu d\theta^2 +ds^2  $. After having defined the tangent vector along the null-path, $\displaystyle V^{\mu} = \frac{dx^\mu}{d\hat{\lambda}} $, it should satisfy the condition $  \mu \left( \frac{d\theta}{d\hat{\lambda}}\right) ^2 +\left( \frac{ds}{d\hat{\lambda}}\right) ^2= 0 $.

The Hamilton equations are equivalent to the Euler-Lagrange equations:
\begin{equation}\label{eom-app1}
\frac{d}{d\hat\lambda} \frac{\partial L}{\partial\left(  dx^{\bar{\mu}} / d\hat\lambda\right) }-
\frac{\partial L}{\partial x^{\bar{\mu}}}= 0\; .
\end{equation}
We now skip some technical details, because a similar derivation is proposed in appendix \ref{app2}, discussing de Broglie's approach. The equation for the fifth component is a conservation law, because the tensor $ a_{\bar{\mu}\bar{\nu}} $ does not depend on the fifth coordinate $ x^5 $. The conserved momentum $ p_5 $ reads $\displaystyle p_5 = \frac{\partial L}{\partial\left(  dx^5 / d\hat\lambda\right) } = \mu\frac{d\theta}{d\hat{\lambda}} $. This conservation law can be used to reduce equations (\ref{eom-app1}), with $ \bar{\mu} = 0,1,2,3 $, to:
\begin{equation}\label{Lorentz-Klein}
mc\left( \frac{d}{d\hat{\lambda}}\left( g_{\mu\nu}V^{\nu}\right) -\frac{1}{2}\partial_\mu g_{\rho\nu}V^\rho V^\nu\right) =-\frac{e}{c}\left( \partial_\mu A_\nu-\partial_{\nu}A_{\mu}\right) V^\nu\, .
\end{equation} 
Klein now introduces the particle's proper time $ \tau $ as follows. The constancy of $ p_5 $ and the condition for the null-like character of the paths imply that the ratio $ \displaystyle \frac{d\tau}{d\hat{\lambda}} $ is constant along the path. Hence, in the projected four-dimensional equation (\ref{Lorentz-Klein}), the arbitrary parameter can be substituted with the proper time, notwithstanding we started considering null-geodesics\footnote{It is worth remembering that the proper-time cannot be defined for null-geodesics.}. After some manipulation it can be shown that it is equivalent to the Lorentz equation for a charged massive particle of mass $ m $ and charge $ -e $ in a combined electromagnetic and gravitational field ( see appendix \ref{app2}): 
\begin{equation}\label{Lorentz}
mc\left( \frac{du^{\lambda}}{d\tau} +\Gamma^\lambda_{\varrho\nu} u^\rho u^\nu\right) =-\frac{e}{c}{F^\lambda}_{\nu} u^\nu\, ,
\end{equation}
where now $\displaystyle u^{\mu} = \frac{dx^\mu}{d\tau} $
is the particle's four-velocity. We stress again the role of the tensor $ a_{\bar{\mu}\bar{\nu} }$. The mass of the particle is hidden into its definition, equation (\ref{metric-K1}). Therefore, the five-dimensional null-geodesics for the differential form $ a_{\bar{\mu}\bar{\nu}}dx^{\bar{\mu}}dx^{\bar{\nu}} $ are connected with four-dimensional geodesics of charged massive particles.

\subsection{{On the inconsistency of a time-like compactified dimension}}\label{app1}
One of the most important assertion we made in the text is that, unlike Klein, de Broglie considered a time-like fifth dimension. In order to understand the consequences of this choice we start again with the five-dimensional line element $ d\sigma^2 = \gamma_{\bar{\mu}\bar{\nu}}dx^{\bar{\mu}}dx^{\bar{\nu}} $. Using Klein notation, which we rewrite here for convenience, we define $ \frac{\gamma_{5\mu}}{\alpha}=\beta A_\mu $ and the components of the five-dimensional metric are:
\begin{equation}\label{metrica}
\gamma_{\mu\nu} = g_{\mu\nu}+\alpha\beta^2A_\mu A_\nu\; , \qquad 
\gamma_{55} = \alpha\; ,\qquad
\gamma_{5\mu} = \alpha\beta A_\mu\; .
\end{equation}
This metric has the following signature: $ \left( -\, ;+\, ;+\, ;+\, ;\epsilon\right)  $, where $ \epsilon = + $ if $ \alpha > 0 $, i.e. if the fifth coordinate describes a space-like dimension, and $ \epsilon = - $ if $ \alpha < 0 $, i.e. in the case of a time-like coordinate. We remember that the line element can be rewritten as $ d\sigma^2 = \alpha d\theta^2 +ds^2 $, where $ d\theta = dx^5 + \beta A_\mu dx^\mu $ and $ ds^2 = g_{\mu\nu}dx^\mu dx^\nu $. The components of the inverse metric are:
\begin{equation}\label{metrica-inversa}
\gamma^{\mu\nu} = g^{\mu\nu}\; , \qquad
\gamma^{55} = \frac{1}{\alpha}+\beta^{2}A_\mu A^\mu\; ,\qquad
\gamma^{5\mu} = -\beta A^\mu\; .
\end{equation}
Using the Ansatz that the metric does not depend on the fifth coordinate, we have calculated the components of the five-dimensional Ricci tensor, defined by
\begin{equation}\label{def-R-tilde}
\tilde{R}_{\bar{\mu}\bar{\nu}}=
\partial_{\bar{\lambda}}\tilde{\Gamma}_{\bar{\mu}\bar{\nu}}^{\bar{\lambda}} -
\partial_{\bar{\nu}}\tilde{\Gamma}_{\bar{\mu}\bar{\lambda}}^{\bar{\lambda}} +
\tilde{\Gamma}_{\bar{\sigma}\bar{\lambda}}^{\bar{\lambda}}\tilde{\Gamma}_{\bar{\mu}\bar{\nu}}^{\bar{\sigma}} -
\tilde{\Gamma}_{\bar{\sigma}\bar{\nu}}^{\bar{\lambda}}\tilde{\Gamma}_{\bar{\mu}\bar{\lambda}}^{\bar{\sigma}}
\; .
\end{equation}
We need the following results:
\begin{eqnarray}
\tilde{R}_{55} &=& \frac{\alpha^{2}\beta^{2}}{4}F_{\mu\nu}F^{\mu\nu}\; ,\label{R1}\\
\tilde{R}_{5\sigma} &=& \alpha\beta\nabla_\lambda{F_\sigma}^{\lambda}+\frac{\alpha^{2}\beta^{3}}{4}A_{\sigma}F_{\mu\nu}F^{\mu\nu}\; ,\label{R2}\\
g^{\mu\nu}\tilde{R}_{\mu\nu} &=& R + \frac{\alpha^{2}\beta^{4}}{4}A_\sigma A^\sigma F_{\mu\nu}F^{\mu\nu} -\frac{\alpha\beta^{2}}{2}F_{\mu\nu}F^{\mu\nu} +
\alpha\beta^{2}A_\mu\nabla_\lambda F^{\mu\lambda}\; ,\label{R3}
\end{eqnarray}
that lead to the following relation for the five-dimensional curvature scalar:
\begin{eqnarray}\label{curv-scal-5}
\tilde{R} &=& \gamma^{\bar{\mu}\bar{\nu}}\tilde{R}_{\bar{\mu}\bar{\nu}} = 
\gamma^{55}\tilde{R}_{55}+2\gamma^{5\mu}\tilde{R}_{5\mu} + \gamma^{\mu\nu}\tilde{R}_{\mu\nu} \nonumber\\
&=& R - \frac{\alpha\beta^{2}}{4}F_{\mu\nu}F^{\mu\nu}\; .
\end{eqnarray}
Equation (\ref{curv-scal-5}) shows that if the fifth dimension is space-like, $ \alpha > 0 $, we can identify $ \alpha\beta^{2} = 2\kappa $ and the electromagnetic kinetic term has the correct sign. On the contrary,  if $ \alpha $ is negative this identification is not possible. This is the inconsistency connected with a compactified time-like dimension. As written in the main text, Klein inferred from this fact the need to introduce a space-like compact dimension.

\subsection{{Geodesics in de Broglie-Rosenfeld approach}}\label{app2}
In this section we describe de Broglie's analysis of five-dimensional geodesics, with some details. After having introduced the five-dimensional metric, in the fifth paragraph of his paper de Broglie considered all five-dimensional geodesics, not only null-geodesics as suggested by Klein, with the following motivation: `Admitting the existence of a fifth dimension of the Universe, we could enunciate the following principle: <<\textit{In the five-dimensional universe, the World-line of every point particle is a geodesic}>>' (\cite{deBroglie}; p. 69). Given $ O $ and $ M $, `two fixed points of the World-line' (\cite{deBroglie}; p. 69), five-dimensional geodesics can be seen as world-lines of extremal ``five-dimensional proper time'' $ d\hat{\tau}=\sqrt{-d\sigma^{2}} $:
\begin{equation}
\delta\int_{O}^{M} d\hat{\tau} =0	\, .
\end{equation}
After introducing an arbitrary parameter $ \hat{\lambda} $, the geodesic equation can be obtained equivalently by the following variational principle:
\begin{eqnarray}
\frac{1}{2}\delta\int_{O}^{M} \left[ \gamma_{\bar{\mu}\bar{\nu}} \frac{dx^{\hat{\mu}}}{d\hat{\lambda}}\frac{dx^{\hat{\nu}}}{d\hat{\lambda}}\right]  \, d\hat{\lambda} =\frac{1}{2}\delta\int_{O}^{M} \left[  \alpha \left( \frac{d\theta}{d\hat\lambda} \right)^2 +g_{\mu\nu}\frac{dx^{\mu}}{d\hat{\lambda}}\frac{dx^{\nu}}{d\hat{\lambda}} \right] d\hat{\lambda} = 0\quad &\text{i.e.}&\nonumber\\
\frac{1}{2}\delta\int_{O}^{M} \left[  \alpha \left( V^5+\beta A_\mu V^\mu\right)^2 +g_{\mu\nu}V^{\mu}V^{\nu} \right] d\hat{\lambda} &=& 0\; ,
\end{eqnarray}
where we used $ d\sigma^2 = \gamma_{\hat{\mu}\hat{\nu}}dx^{\hat{\mu}}dx^{\hat{\nu}} = \alpha d\theta^2 +ds^2 $ and where $ V^5 $ and $ V^\mu $ are the five components of the five-velocity $ \displaystyle V^{\bar\mu}=\frac{dx^{\bar\mu}}{d\hat\lambda} $. Now de Broglie identified the quantity into the square bracket as a Lagrangian $ L(x\, ,\, V) $. Varying the action as a function of $ x^{\bar\mu} $ and $ V^{\bar\mu} $, de Broglie obtained the following Euler-Lagrange equations: 
\begin{subequations}
\begin{eqnarray} 
\frac{d}{d\hat{\lambda}}\frac{\partial L}{\partial V^5} &=& \frac{\partial L}{\partial x^5}\; , \label{eq-moto1} \\
\frac{d}{d\hat{\lambda}}\frac{\partial L}{\partial V^\mu  } &=& \frac{\partial L}{\partial x^{\mu}}\label{eq-moto2}\; .
\end{eqnarray}
\end{subequations}
Remembering that there is no dependence from the fifth dimension, the equation (\ref{eq-moto1}) produces a conserved quantity:
\begin{equation}\label{p5}
\frac{d}{d\hat{\lambda}}\alpha\left(  V^5+\beta A_{\mu}V^{\mu}\right)  = 0\qquad \text{i.e.} \qquad \pi_5=  \alpha \frac{d\theta}{d\hat\lambda} = \text{constant}\; ,
\end{equation}
while equations (\ref{eq-moto2}) read\footnote{Remember that $ A_\mu $ is a function of the four-dimensional coordinates.}:
\begin{equation}
\frac{d}{d\hat\lambda}\left( \pi_5\beta A_\mu+g_{\mu\nu}V^\nu\right)
=  \frac{1}{2}\partial_\mu g_{\rho\sigma}V^\rho V^\sigma + \pi_5\beta 
\partial_\mu A_\nu V^\nu  \; ,
\end{equation}
and, rearranging the terms and inserting $ \pi_5 $ expression (\ref{p5}), its equivalent form is:
\begin{equation}\label{eq-moto3}
\frac{d}{d\hat\lambda}\left( g_{\mu\nu}\frac{dx^{\nu}}{d\hat{\lambda}}\right) =
\frac{1}{2}\partial_\mu g_{\rho\sigma}\frac{dx^{\rho}}{d\hat{\lambda}} \frac{dx^{\sigma}}{d\hat{\lambda}} + \alpha \frac{d\theta}{d\hat\lambda}
\beta F_{\mu\rho}\frac{dx^{\rho}}{d\hat{\lambda}}\, .
\end{equation}
We can now introduce the proper-time $ d\tau = \sqrt{-ds^2} $, because we are considering non-null geodesics. The five-dimensional geodesic equation and the metricity condition imply that the covariant derivative of the $ \gamma_{\bar{\mu}\bar{\nu}}V^{\bar{\mu}}V^{\bar{\nu}} $ would be zero. Hence the ratio $ \frac{d\hat{\lambda}}{d\tau} $ is constant along the geodesic curve and in equation (\ref{eq-moto3}) $ \hat{\lambda} $ could be substituted by $ \tau $. If we define the normalized four-dimensional vector $ \displaystyle u^{\mu} = \frac{dx^{\mu}}{d\tau} $ and if we set, following de Broglie,
\begin{equation}\label{rel1}
\alpha\frac{d\theta}{d\tau}= -\frac{e}{\beta c}\dfrac{1}{mc}\; ,
\end{equation}
equation (\ref{eq-moto3}) reduces to
\begin{equation}\label{eq-lorentz2} 
mc\left( \frac{d}{d\tau}\left( g_{\mu\nu}u^{\nu}\right) -\frac{1}{2}\partial_\mu g_{\rho\nu}u^\rho u^\nu\right) =-\frac{e}{c}F_{\mu\nu} u^\nu\, .
\end{equation}
As claimed in the main text, the parameter $ \beta $ disappears and it remains undetermined.

In order to obtain Lorentz equations we rewrite the first term of the l.h.s. of equation (\ref{eq-lorentz2}) as follows:
\begin{equation}
 \frac{d}{d\tau}\left(  g_{\mu\nu}u^{\nu}\right) =
 u^{\rho}\partial_{\rho} 
 \left( g_{\mu\nu }u^{\nu} \right) = g_{\mu\nu}
 \frac{du^{\nu}}{d\tau}+ \frac{1}{2}\left( \partial_{\rho} g_{\mu\nu}+\partial_{\nu} g_{\mu\rho} \right) \; ,
\end{equation}
Finally, we insert the previous equation in (\ref{eq-lorentz2}) and we contract it with the inverse components of the metric $ g^{\lambda\mu} $ to get the Lorentz equations:
\begin{equation}\label{eq-lorentz3} 
mc\left( \frac{du^{\lambda}}{d\tau} +\Gamma^{\lambda}_{\mu\rho}u^\rho u^\nu\right) =-\frac{e}{c}{F^{\lambda}}_{\nu} u^\nu\, .
\end{equation}

Unlike Klein, de Broglie's purpose was to show how the five-dimensional Universe's approach permits to geometrize the electromagnetic force. He stressed: `This means that with geometric meaning we have attributed to the [electromagnetic] potentials and to the ratio $ e/m $, the five-dimensional World-lines of point particles are all geodesics. \textit{The notion of force has been completely banned from Mechanics}.' (\cite{deBroglie}; p. 70). This connection between geodesic lines and equation (\ref{eq-lorentz2}) convinced de Broglie that was not necessary to consider null-geodesics lines only.

De Broglie's investigation of five-dimensional geodesic lines continued with the question of what would be the correct particle's action in five dimensions. The author proposed (\cite{deBroglie}; p. 70):
\begin{equation}\label{azione-dB}
S_5 = -\mathcal{I} \int_O^M d\hat\tau \, ,
\end{equation}
where
\begin{equation}
	\mathcal{I} = m^{2}c^{2}-\frac{e^{2}}{\alpha\beta^{2}c^{2}}\; ,
\end{equation}
because it reduces to the usual point particle action in the case of zero charge. In order to understand this fact,  following de Broglie, we point out that we can rewrite $ S_5 $ as follows:
\begin{equation}\label{action-final1}
S_5 = -\mathcal{I} \int_O^M d\hat\tau = \int_O^M\left(\mathcal{I}\alpha\frac{d\theta}{d\hat{\tau}} \right) d\theta -\int_O^M \left( \mathcal{I}\frac{d\tau}{d\hat{\tau}}\right) d\tau \; , 
\end{equation}
where the second equality sign follows by inserting $\displaystyle 1 = \left( \frac{d\hat{\tau}}{d\hat{\tau}}\right) ^2 = \left( \frac{d\tau}{d\hat{\tau}}\right) ^2 - \alpha \left( \frac{d\theta}{d\hat{\tau}}\right) ^2  $, as a formal consequence of $d\hat{\tau}^2 = d\tau^2 - \alpha d\theta^2  $. We remember that the condition $ \partial_5\gamma_{\bar{\mu}\bar{\nu}}=0 $ is equivalent to assert, using modern language, that space-time would admit a Killing vector field, which is tangent to the fifth coordinate. The scalar product between the Killing field and the velocity field is constant along the geodesic. This result implies that the ratio $ \frac{d\theta}{d\hat{\tau}} $ must be constant. Hence, de Broglie chose:
\begin{equation}\label{rel2}
\mathcal{I}\frac{d\tau}{d\hat{\tau}} = mc
\end{equation}
and
\begin{equation}\label{rel1b}
\mathcal{I}\alpha\frac{d\theta}{d\hat{\tau}} = -\frac{e}{c\beta}\; ,
\end{equation}
which are consistent with equations (\ref{rel1}).
Finally, using $ d\theta = dx^5 + \beta A_\mu dx^\mu $, $ S_5 $ assumes the following form:
\begin{equation}\label{actionS5}
 S_5=-\int \frac{e}{c\beta}dx^5+ \frac{e}{c}\int A_\mu dx^{\mu} -mc\int d\tau \; .
\end{equation}
It is now evident that $ S_5 $ reduces to $\displaystyle S_4=-mc\int  d\tau $ when we set $ e=0 $. Indeed, when $ e=0 $ the scalar product between the Killing field and the velocity field (\ref{rel1b}) (cf. de Broglie's comment on equation (\ref{projection})) is zero, then the geodesic projects onto the four-dimensional space-time. As a consequence, de Broglie convinced himself that the invariant $ \mathcal{I}^{2} $ should have been a five-dimensional generalization of the four-dimensional momentum\footnote{We remember that De Broglie's idea emerged comparing $ S_4 $ with $ S_5 $.} $ mc $. At this stage, we are able to explain the form of the invariant $ \mathcal{I} $. Equations (\ref{rel2}) and (\ref{rel1b}) and the identity $ d\hat{\tau}^2 = d\tau^2 \alpha d\theta^2 $ imply that $ \mathcal{I} $ must have the following form: 
\begin{equation}\label{formaI}
\mathcal{I}^2 = m^2c^2 - \frac{e^{2}}{\alpha\beta^{2}c^{2}} \; .
\end{equation}

As noted by Klein, de Broglie choose $ -\alpha\beta^{2} = 2\kappa $, because, if the fifth dimension is time-like, $ \alpha $ is negative and the invariant $ \mathcal{I}^{2} $ would be strictly positive. On the other hand, as we have said, the choice is not consistent with the request to obtain Maxwell Lagrangian in (\ref{curv-scal-5}). 

As we have said in the main text, Klein asserted that de Broglie's mistake did not affect his conclusions. Klein referred to the following fact.
De Broglie proposed the five-dimensional wave equation:
\begin{equation}\label{we}
\gamma^{\bar{\mu}\bar{\nu}}\nabla_{\bar{\mu}}\partial_{\bar{\nu}}\Psi\,=\,\frac{4\pi^2}{h^2} \mathcal{I}^2\Psi \; .
\end{equation}
It is worth noting that $ S_5 $ depends linearly from $ x^5 $, as we can see integrating (\ref{actionS5}). Hence, using a geometrical optics Ansatz $ \displaystyle\Psi=Ae^{\frac{i}{\hbar}S_5} $, the periodicity with respect to $ x^5 $ follows immediately.
This means that the wave function can be written as:
\begin{equation}\label{psi1}
\Psi\left( x\right)  = \psi\left( x^0\, ,x^1\, , x^2\, , x^3 \right)\cdot \text{exp}\left( \frac{i}{\hbar}\frac{e}{c\beta}x^5\right) \; ,
\end{equation}
where $ \psi $ is the four-dimensional wave function. Using (\ref{psi1}) and the components of the inverse metric (\ref{metrica-inversa}), we note, following Klein (\cite{Klein-risp-deBroglie}; p. 243) that, $ \Psi $ satisfies 
\begin{equation}\label{rel3}
\gamma^{55}\partial^2_5\Psi = -\frac{1}{\hbar^{2}}\left( 
\frac{1}{\alpha}+\beta^{2}A_{\mu}A^{\mu} \right) \frac{e^2}{c^2\beta^2}\Psi\; .
\end{equation}
This means that (\ref{we}) can be rewritten, in a flat space-time, in the following way (\cite{deBroglie}; p. 73):
\begin{equation}\label{ultima1}
g^{\mu\nu}\partial_\mu \partial_\nu\psi - \frac{2ie}{\hbar c}A^\mu\partial_\mu\psi - \frac{e^2}{\hbar^2 c^2}A_\mu A^\mu\psi = \frac{m^2c^2}{\hbar^2}\psi\; .
\end{equation}
We note that (\ref{ultima1}) corresponds to the KG equation for a complex scalar field in an external electromagnetic field, and can be written in the following compact way:
\begin{equation}
g^{\mu\nu}\left(\partial_{\mu}-\frac{i}{\hbar}\frac{e}{c}A_{\mu} \right)\left(\partial_{\nu}-\frac{i}{\hbar}\frac{e}{c}A_{\nu} \right)\psi = \frac{m^2c^2}{\hbar^2}\psi  \; , 
\end{equation}
if the Lorenz gauge, namely $\displaystyle\partial_{\mu} A^{\mu}=0 $, holds. As claimed by Klein, independently to the character of the fifth dimension, the term depending on $ \alpha\beta^{2} $ in $ \mathcal{I}^{2} $ definition (\ref{formaI}) disappears, and equation (\ref{ultima1}) reduces to de Broglie's equation (\cite{deBroglie}; p. 73, equation (40)), where the case of null gravitational field is considered.

\subsection{{On Rosenfeld approach}}
In this section we explain some technical details skipped in the main text. 

\subsubsection{{Five-dimensional versus four-dimensional momentum}}\label{app3a}
Equations (\ref{RosB}) and (\ref{slope}) can be obtained as follows. First we note that if $ S_0 $ is a complete integral of the HJ equation in four dimensions, see equation (\ref{HJ-Ros1}), it follows that\footnote{See \cite{Landau-teo-campi}.} $ \displaystyle g^{\mu\nu}\left( \partial_\nu S_0 + \frac{e}{c}A_\nu \right) = mc \frac{dx^\mu}{d\tau} $. Then, using the inverse components of the metric tensor (\ref{inverse}) and equation (\ref{eq7}), we rewrite the l.h.s. of (\ref{RosB}) as follows:
\begin{eqnarray}
\gamma^{\mu\bar{\nu}}\partial_{\bar{\nu}}\bar{S} &=& \gamma^{\mu 5}\partial_{5}\bar{S} + \gamma^{\mu\nu}\partial_{\nu}\bar{S} = \left( -\beta A^{\mu}\right) \left( -\frac{e}{c\beta}\right) + g^{\mu\nu}\partial_\nu S_0\nonumber\\
&=& g^{\mu\nu}\left( \partial_\nu S_0 +\frac{e}{c}A_{\nu}\right) =  mc \frac{dx^\mu}{d\tau}\, ,
\end{eqnarray}
and we have finally obtained equation  (\ref{RosB}). Since Rosenfeld introduced explicitly the quantity $ \sqrt{m^2c^2 - \frac{e^2c^2}{16\pi G}} $, we used for this quantity the symbol $ \mathcal{I}_{Ros} $ for brevity. From equation (\ref{RosA}) we get
\begin{equation}\label{RosA1}
\gamma^{\mu\bar{\nu}}\partial_{\bar{\nu}}\bar{S}=\mathcal{I}_{Ros}\frac{dx^{\mu}}{d\hat{\tau}} = \mathcal{I}_{Ros}\frac{dx^{\mu}}{d\tau}\frac{d\tau}{d\hat{\tau}} \, ,
\end{equation}
and confronting equation (\ref{RosA1}) with  (\ref{RosB}) we get equation (\ref{slope}).

\subsubsection{{Modern five-dimensional action}}\label{app3b}
In action (\ref{az-glob2}) Rosenfeld choose an unusual coupling between matter and gravity. Rosenfeld's coupling is unusual for the following reason. In a modern five-dimensional approach, the action would be:
\begin{equation}\label{az-reale}
\mathcal{S}_{tot} \left( \gamma\, ,\,\Phi\, ,\,\bar{\Phi}\right) = 
\int d^5x\sqrt{-\gamma}\left[ -\dfrac{1}{2\kappa_5}\tilde{R}+\tilde{\mathcal{L}}\right] \; , 
\end{equation}  
where $ \tilde{\mathcal{L}} $ is the action for a complex scalar field $ \Phi $, that has the expected length dimension $ \left[ \Phi\right] = (length)^{-\frac{3}{2}}  $, in natural units $ \hbar = c = 1 $.
Using the determinant definition and (\ref{metrica}) it can be proved that\footnote{We define $ \epsilon^{01235} = 1 $.}
\begin{equation}
\gamma = \epsilon^{\bar{\mu}\bar{\nu}\bar{\rho}\bar{\sigma}\bar{\lambda}}\gamma_{\bar{\mu}0}
\gamma_{\bar{\nu}1}\gamma_{\bar{\rho}2}\gamma_{\bar{\sigma}3}\gamma_{\bar{\lambda}5} = \alpha g\; .
\end{equation}
Using $ \kappa_5 = 2\pi \tilde{l}\kappa$, where $ 2\pi \tilde{l} $ is the ``volume'' of the compact dimension, (\ref{az-reale}) can be rewritten as follows:
\begin{equation}\label{az-reale2}
\mathcal{S}_{tot} \left( \gamma\, ,\,\Phi\, ,\,\bar{\Phi}\right) = 
\dfrac{\sqrt{\alpha}}{2\kappa_5}\int d^5x\sqrt{-g}\left[ -\tilde{R}+\kappa \left( 2\pi\tilde{l}\tilde{\mathcal{L}}\right) \right] \; . 
\end{equation}
Now the length $ \tilde{l} $ of the fifth dimensions can be adsorbed with the following field redefinition: $ \Psi = \sqrt{2\pi\tilde{l}}\,\Phi $. This shows that the equations obtained by varying (\ref{az-reale2}) are equivalent to Rosenfeld's equations of motion, but the new scalar field $ \Psi $ has length dimensions $ \left[ \Psi\right] = (length)^{-1} $ as a four-dimensional scalar field. As a consequence, as stated in the main text, the stress-energy tensor defined by Rosenfeld is a four-dimensional object. 

\subsubsection{{Einstein-Maxwell equations coupled with complex scalar field}}\label{app3c}
The equations obtained by varying (\ref{az-reale2}) with respect to the metric are:
\begin{equation}\label{eomA}
\tilde{R}_{\bar{\mu}\bar{\nu}}-\frac{1}{2}\gamma_{\bar{\mu}\bar{\nu}}\tilde{R} = \kappa T_{\bar{\mu}\bar{\nu}} \; ,
\end{equation}
and, as written in the main text, they are formally equivalent to the the four-dimensional Einstein equations, coupled to the electromagnetic and matter stress-energy tensor, and Maxwell equations. In order to understand this fact, firstly we write the expression for $ \tilde{R}_{\mu\nu} $. After a lengthy calculation, from (\ref{def-R-tilde}) it follows:
\begin{equation}\label{R3b}
\tilde{R}_{\mu\nu} = R_{\mu\nu} + \frac{\alpha^{2}\beta^{4}}{4}A_\mu A_\nu F_{\sigma\lambda}F^{\sigma\lambda} -\frac{\alpha\beta^{2}}{2}F_{\mu\lambda}{F_\nu}^{\lambda} +
\frac{\alpha\beta^{2}}{2}\left( A_\mu\nabla_\lambda {F_\nu}^{\lambda} + A_\nu\nabla_\lambda {F_\mu}^{\lambda}\right) \; .
\end{equation}
Let us consider the contravariant components of (\ref{eomA}), i.e.
\begin{equation}\label{eomB}
\gamma^{\bar{\lambda}\bar{\mu}}\gamma^{\bar{\sigma}\bar{\nu}}\tilde{R}_{\bar{\mu}\bar{\nu}}-\frac{1}{2}\gamma^{\bar{\lambda}\bar{\sigma}}\tilde{R} = \kappa\gamma^{\bar{\lambda}\bar{\mu}}\gamma^{\bar{\sigma}\bar{\nu}}T_{\bar{\mu}\bar{\nu}}\; .
\end{equation}
Using (\ref{metrica}), (\ref{metrica-inversa}), (\ref{R1}), (\ref{R2}) and (\ref{R3b}) the $ \lambda\sigma $ components of the l.h.s. of equation (\ref{eomB}) can be rewritten as follows:
\begin{eqnarray}
\gamma^{\lambda\bar{\mu}}\gamma^{\sigma\bar{\nu}}\tilde{R}_{\bar{\mu}\bar{\nu}}-\frac{1}{2}\gamma^{\lambda\sigma}\tilde{R} &=& \left[ g^{\lambda\mu}g^{\sigma\nu}\tilde{R}_{\mu\nu}+g^{\lambda\mu}\gamma^{\sigma 5}\tilde{R}_{5\mu} + 
\gamma^{\lambda 5}g^{\sigma \nu}\tilde{R}_{5\nu} + \gamma^{\lambda 5}\gamma^{\sigma 5}\tilde{R}_{55} \right] -\frac{1}{2}g^{\lambda\sigma}\tilde{R}\, , \nonumber\\
&=&  R^{\lambda\sigma}-\frac{1}{2}g^{\lambda\sigma}R - \kappa g^{\lambda\mu}g^{\sigma\nu} T^{em}_{\mu\nu}\; .
\end{eqnarray}
Following Rosenfeld we define
\begin{equation}
T^{\lambda\sigma}=\gamma^{\lambda\bar{\mu}}\gamma^{\sigma\bar{\nu}}T_{\bar{\mu}\bar{\nu}}\; ,
\end{equation}
and the $ \lambda\sigma $ components of (\ref{eomB}) read (\cite{Ros1}; p. 313):
\begin{equation}\label{Einstein}
R^{\lambda\sigma}-\frac{1}{2}g^{\lambda\sigma}R = \kappa \left(  T_{em}^{\lambda\sigma} + T^{\lambda\sigma} \right) \; ,
\end{equation}
that correspond to Einstein equations coupled to the electromagnetic and the matter stress-energy tensor. Maxwell equations emerge conversely as follows. If we contract (\ref{eomA}) with $ \gamma^{\bar{\rho}\bar{\mu}} $, we get:
\begin{equation}\label{eomC}
\gamma^{\bar{\rho}\bar{\mu}}\tilde{R}_{\bar{\mu}\bar{\nu}}-\frac{1}{2}{\delta^{\bar{\rho}}}_{\bar{\nu}}\tilde{R} = \kappa\gamma^{\bar{\rho}\bar{\mu}} T_{\bar{\mu}\bar{\nu}}\; .
\end{equation}
The $ \rho 5 $ components of the l.h.s. of equation (\ref{eomC}) now read\footnote{Remember that $ {\delta^{\bar{\rho}}}_{\bar{\nu}}=0 $ when $ \bar{\rho}\neq \bar{\nu}$}:
\begin{eqnarray}
\gamma^{\rho\bar{\mu}}\tilde{R}_{\bar{\mu}\bar{\nu}} &=& \gamma^{\rho\mu}\tilde{R}_{\mu 5} +
\gamma^{\rho 5}\tilde{R}_{5 5}\; ,\nonumber\\
&=& \frac{\alpha\beta}{2}\nabla_\lambda F^{\rho\lambda}\; .
\end{eqnarray}
Remembering that $ \kappa = \frac{\alpha\beta^2}{2} $, following Rosenfeld, we define
\begin{equation}
{T^\rho}_5 = \gamma^{\rho\bar{\mu}} T_{\bar{\mu}5}\; ,
\end{equation}
and equation (\ref{eomC}) now reads:
\begin{equation}\label{em}
\nabla_\lambda F^{\rho\lambda} = \beta {T^\rho}_5 \; .
\end{equation}
Equations (\ref{em}) correspond to Maxwell equations coupled to a current density as written by Rosenfeld (\cite{Ros1}; p. 313).

\subsubsection{{Four-dimensional and five-dimensional curvature scalar}}\label{app3d}
In the main text, we have written that using (\ref{eomA}) Rosenfeld obtained a particular relation for the curvature scalars $ R $ and $ \tilde{R} $, namely
\begin{eqnarray}
R &=&-\kappa\left[ \gamma^{\nu\bar{\mu}}T_{\bar{\mu}\nu} -\gamma^{\mu\rho}A_\rho\nabla_\lambda\left( \gamma_{\mu\sigma}F^{\sigma\lambda}\right) \right]\qquad \text{and}\label{formula1}\\
\tilde{R} &=& -\kappa\left[ \gamma^{\nu\bar{\mu}}T_{\bar{\mu}\nu}+\frac{F_{\sigma\lambda}F^{\sigma\lambda}}{2} -\gamma^{\mu\rho}A_\rho\nabla_\lambda\left( \gamma_{\mu\sigma}F^{\sigma\lambda}\right) \right].\label{formula2} 
\end{eqnarray}
In order to obtain these relations, we set $ \bar{\rho}=\bar{\nu}=\nu $ in equation (\ref{eomC}) and it reads:
\begin{equation}\label{eomD}
\gamma^{\nu\bar{\mu}}\tilde{R}_{\bar{\mu}\nu}-2\tilde{R} = \kappa {T^{\nu}}_\nu\; ,
\end{equation}
where we have defined $\displaystyle {T^{\nu}}_\nu = \gamma^{\nu\bar{\mu}} T_{\bar{\mu}\nu} $. Using the definition of $ \tilde{R} $ (equation (\ref{curv-scal-5})), equation (\ref{eomD}) can be rewritten as
\begin{equation}\label{eomD1}
\tilde{R}-\gamma^{55}\tilde{R}_{55}-\gamma^{5\mu}\tilde{R}_{\mu 5}-2\tilde{R} = \kappa {T^{\nu}}_\nu\; .
\end{equation}
Inserting (\ref{metrica-inversa}), (\ref{R1}), (\ref{R2}) and (\ref{curv-scal-5}), and isolating $ R $, we obtain equation (\ref{formula1}) and using again (\ref{curv-scal-5}) we obtain (\ref{formula2}).

\subsubsection{{The retarded potentials}}\label{app3e}
After having linearised Einstein equations (\ref{eomA}), Rosenfeld integrated it and obtained the retarded potentials, equation (\ref{sol-eq}). Using modern notation the retarded potentials read:
\begin{equation}
h_{\bar{\mu}\nu}(t; \mathbf{x}) = -\frac{\kappa}{2\pi}\int_\Sigma \bar{T}_{\bar{\mu}\nu}\left( t-\frac{| \mathbf{x}- \mathbf{y}|}{c}; \mathbf{y}\right) \frac{d^3y}{| \mathbf{x}- \mathbf{y}|} \quad ,
\end{equation} 
where the radial distance is defined by $ r= | \mathbf{x}- \mathbf{y}|$ and the integration is carried on a three-dimensional hypersurface $ \Sigma $ at the retarded time $\displaystyle t-\frac{| \mathbf{x}- \mathbf{y}|}{c}  $. The retarded potential are functions of $  \mathbf{x} $ and $ t $.

\subsubsection{{The isotropic coordinate system and the ``mean distance''}}\label{app3f}
In this last appendix we show how Rosenfeld was inspired by his knowledge of Eddington's book on GR. 

Given a bounded charged matter distribution of radius $ \epsilon $, the RN metric is an exact solution of equations (\ref{Einstein}), with $ T^{\lambda\sigma} $ being the stress-energy tensor associated to the classical spherical symmetric mass distribution. In polar coordinates the line element has the following form:
\begin{equation}\label{RN}
ds^2 = -\left( 1- \frac{2mG}{c^2r} +\frac{GQ^2}{c^4r^2} \right)c^2dt^2 + \left( 1- \frac{2mG}{c^2r}+\frac{GQ^2}{c^4r^2}\right)^{-1}dr^2 + r^2\left( d\theta^2 + sin^2\theta d\varphi^2\right) \, ,
\end{equation}
where $ m $ and $ Q $ are the mass and the charge of the particle respectively and the coordinate $ r $ has the following range: $ \epsilon\leq r<+\infty $. If $ Q=0 $ the line element describes the so called exterior Schwartzschild metric. Rosenfeld used the less known isotropic coordinate system. We do not know if the author would know RN metric in isotropic coordinates. As stated in section \ref{prehist}, we know from Kuhn's interview \cite{Kuhn1} that Rosenfeld studied Eddington's book on GR. In \textit{The Mathematical Theory of Relativity} \cite{Eddington} the British Physicist introduced isotropic coordinates for Schwartzschild metric, using both its exact form and its limit at first order in $ \displaystyle \frac{1}{r} $. It is worth noting that at asymptotically large distances from the source, at the first order in $ \displaystyle \frac{1}{r} $, both Schwartzschild and RN metric have the same form. This fact is true both in isotropic and in polar coordinates. 

In the so called isotropic Cartesian coordinate system the line element of a spherically symmetric space-time has the following form:
\begin{equation}\label{isotropic1}
ds^2 = -A\left( r\right) dt^2 + B\left( r\right) \left( dx^2+dy^2+dz^2\right)  \, ,
\end{equation}
where $ r=\sqrt{x^2+y^2+z^2} $ is the distance from the origin. Following Eddington, at the first order in $ \displaystyle \frac{1}{r} $, for a point-particle continually at rest we have (\cite{Eddington}; p. 101):
\begin{equation}
A\left( r\right)\approx 1-\frac{2mG}{c^2r}\qquad\text{and}\qquad B\left( R\right) \approx 1+\frac{2mG}{c^2r}\;\, ,
\end{equation}
where the particle need not be at the origin provided that $ r $ is the distance from the particle to the point considered. The line element now reads:
\begin{equation}\label{isotropic2}
ds^2 = -\left( 1-\frac{2mG}{c^2r}\right) dt^2 + \left( 1+\frac{2mG}{c^2r}\right) \left( dx^2+dy^2+dz^2\right)  \, ,
\end{equation}
showing that at large distances the particle's gravitational field is ``less different'' from the Minkowskian field, as stated by Rosenfeld. 

Line element (\ref{isotropic2}) and Rosenfeld's line element are different, see e.g. (\ref{metr-cl-a}). Rosenfeld used the ``mean distance'' $ r_0(\vec{x}) $ instead of $ r $: Rosenfeld replaced the distance to the single particle by the mean distance to the cloud. In order to understand this fact, we remember that inspecting the semi-classical limit of his quantum metric the particle is represented by a wave function that is zero outside a volume $ V $. For this reason, following Eddington, we consider the transition to continuous matter. Summing the fields of force of a number of particles, Eddington suggested the following form for the two functions $ A\left( R\right) $ and $ B\left( R\right) $:
\begin{equation}
A\left( r\right)\approx 1-\frac{2\Omega}{c^2}\qquad\text{and}\qquad B\left( r\right) \approx 1+\frac{2\Omega}{c^2}\;\, ,
\end{equation}
where $ \Omega $ represents the Newton potential at the point considered and using Eddington notation reads:
\begin{equation}\label{particles1}
\Omega = \sum \frac{m}{r}\;\, .
\end{equation}
Let $ \vec{y}_i $, with $ i=1,\dots ,N $, be the position of the i-th particle, $ m_i $ its mass and let $ \vec{x} $ be an arbitrary point of the space-time. Using modern notation, equation (\ref{particles1}) reads:
\begin{equation}\label{particles2}
\Omega = \sum_{i=1}^N \frac{m_i}{|\vec{y}_i-\vec{x}|}\;\, .
\end{equation}
For a homogeneous system of mass $ m $ with volume $ V $ the Newton potential reads:
\begin{equation}\label{homo}
\Omega = \frac{m}{V}\int_{V}\frac{dxdydz}{|\vec{y}-\vec{x}|}\;\, ,
\end{equation}
where $  \mathbf{y} $ is a point of the volume $ V $.
The mean value theorem states that: 
\begin{equation}\label{media}
	\frac{1}{V}\int_V \frac{dxdydz}{| \mathbf{x}- \mathbf{y}|} = \frac{1}{r_0( \mathbf{x})}
\end{equation}
where $ r_0(\vec{x}) $ is the mean distance to the cloud. Equation (\ref{media}) is equivalent to Rosenfeld's condition (\ref{def-r0}), namely $\displaystyle \frac{V}{r_0( \mathbf{x})} =\int_V \frac{dV}{| \mathbf{x}- \mathbf{y}|} $ and the line element to be compared with the semi-classical limit of the quantum metric reads:
\begin{equation}\label{isotropic3}
ds^2 = -\left( 1-\frac{2mG}{c^2r_0(\vec{x})}\right) dt^2 + \left( 1+\frac{2mG}{c^2r_0(\vec{x})}\right) \left( dx^2+dy^2+dz^2\right)  \, .
\end{equation}

\end{document}